\documentclass[twocolumn,showpacs,preprintnumbers,amsmath,amssymb,latexsym,prd,footinbib,floatfix]{revtex4-2}

\usepackage{amssymb,amsmath,bm}

\usepackage[pdftex, pdftitle={Impossibility of spontaneous vector
  flavor symmetry breaking on the lattice}]{hyperref}

\newcommand{\f}[2]{\frac{#1}{#2}}
\newcommand{\tf}[2]{{\textstyle\f{#1}{#2}}}

\newcommand{\la}{\langle}
\newcommand{\ra}{\rangle}

\newcommand{\de}{\partial}

\newcommand{\tr}{{\rm tr}}
\newcommand{\Tr}{{\rm Tr}}
\newcommand{\diag}{\mathrm{diag}}
\newcommand{\act}{\mathcal{S}}

\newcommand{\gmeas}{\mathrm{D}U}
\newcommand{\fmeas}{\mathrm{D}\psi\mathrm{D}\bar{\psi}}

\newcommand{\Dirac}{\mathrm{D}}
\newcommand{\propa}{\mathrm{S}}

\newcommand{\dml}{D^{(0)}}
\newcommand{\dmls}{D^{\mathrm{S}(0)}}
\newcommand{\dmlgw}{D^{\mathrm{GW}(0)}}
\newcommand{\dmlw}{D^{\mathrm{W}(0)}}
\newcommand{\dmln}{D^{\mathrm{n}(0)}}

\newcommand{\dmlmd}{D^{\mathrm{X}(0)}}
\newcommand{\dmlgen}{D^{\mathrm{X}(0)}}

\newcommand{\deltaD}{\Delta D}
\newcommand{\ded}{\Delta D_{\mathrm{max}}}

\newcommand{\colA}{\mathbf{A}}
\newcommand{\colB}{\mathbf{B}}
\newcommand{\colx}{\mathbf{x}}
\newcommand{\coly}{\mathbf{y}}

\newcommand{\Asc}{{A_{\star}}} 
\newcommand{\Bsc}{{B_{\star}}}
\newcommand{\Csc}{{C}}
\newcommand{\Cscc}{{C_{\star}}}
\newcommand{\Ms}{\tilde{\mathcal{M}}}
\newcommand{\Vc}{{\mathcal V}}
\newcommand{\Uc}{{\mathcal U}}
\newcommand{\Tc}{{\mathcal T}}
\newcommand{\Cc}{{\mathcal C}}
\newcommand{\Fc}{{\mathcal F}}
\newcommand{\Mc}{{\mathcal M}}
\newcommand{\Nc}{{\mathcal N}}
\newcommand{\Sc}{{\mathcal S}}
\newcommand{\Kc}{{\mathcal K}}
\newcommand{\Oc}{{\mathcal O}}

\begin{document}

\title{Impossibility of spontaneous vector flavor
  symmetry breaking on the lattice}

\author{Matteo Giordano}
\email{giordano@bodri.elte.hu}
\affiliation{ELTE E\"otv\"os Lor\'and University, Institute for
  Theoretical Physics, P\'azm\'any P\'eter s\'et\'any 1/A, H-1117, Budapest,
  Hungary}

\begin{abstract}
  I show that spontaneous breaking of vector flavor symmetry on the
  lattice is impossible in gauge theories with a positive
  functional-integral measure, for discretized Dirac operators linear
  in the quark masses, if the corresponding propagator and its
  commutator with the flavor symmetry generators can be bounded in
  norm independently of the gauge configuration and uniformly in the
  volume. Under these assumptions, any order parameter vanishes in the
  symmetric limit of fermions of equal masses. I show that these
  assumptions are satisfied by staggered, minimally doubled and
  Ginsparg-Wilson fermions for positive fermion mass, for any value of
  the lattice spacing, and so in the continuum limit if this
  exists. They are instead not satisfied by Wilson fermions, for which
  spontaneous vector flavor symmetry breaking is known to take place
  in the Aoki phase. The existence of regularizations unaffected by
  residual fermion doubling for which the symmetry cannot break
  spontaneously on the lattice establishes rigorously (at the
  physicist's level) the impossibility of its spontaneous breaking in
  the continuum for any number of flavors.
\end{abstract}

\maketitle

\section{Introduction}
\label{sec:intro}

The importance of symmetries and of the way in which they are realized
in quantum field theories can hardly be overemphasized. In the context
of strong interactions and its microscopic theory, i.e., QCD, an
important role is played by the approximate vector flavor symmetry
involving the lightest two or three types (``flavors'') of quarks,
which holds exactly in the limit of quarks of equal masses; and by its
enhancement to chiral flavor symmetry in the limit of massless
quarks. Vector flavor symmetry and the pattern of its explicit
breaking largely determine the structure of the hadronic spectrum;
chiral flavor symmetry and its spontaneous breaking down to vector
flavor symmetry explain the lightness of pions and their dynamics, as
well as the absence of parity partners of hadrons. The full symmetry
group at the classical level includes also the $\mathrm{U}(1)_B$
symmetry responsible for baryon number conservation, and the axial
$\mathrm{U}(1)_A$ symmetry, that does not survive the quantization
process and becomes anomalous in the quantum theory.

An interesting question is whether baryon number and vector flavor
symmetry can break down spontaneously in general vector gauge
theories, where the fermions' left-handed and right-handed chiralities
are coupled in the same way to the gauge fields. This could in
principle happen for exactly degenerate massive fermions, leading to
the appearance of massless Goldstone bosons; and in the chiral limit
of massless fermions it could lead to a different symmetry breaking
pattern than the usual one, and so to a different set of Goldstone
bosons.  This question has been essentially answered in the negative
by Vafa and Witten in a famous paper~\cite{Vafa:1983tf}. There they
actually prove a stronger result, namely the impossibility of finding
massless particles in the spectrum of a gauge theory with positive
functional-integral measure that couple to operators with nonvanishing
baryon number or transforming nontrivially under vector flavor
transformations. This is done by deriving a bound on the fermion
propagator that guarantees its exponential decay with the distance as
long as the fermion mass is nonzero. Since massless bosons coupling to
the operators mentioned above would appear in the spectrum as a
consequence of Goldstone's theorem~\cite{Goldstone:1962es,
  Lange:1965zz,Strocchi:2008gsa} if those symmetries were
spontaneously broken, the impossibility of spontaneous breaking
follows.

The elegant and powerful argument of Vafa and Witten is developed
using the ``mathematical fiction'' of the functional integral
formalism for interacting quantum field theories in continuum
(Euclidean) spacetime. The crucial issue of the regularization of the
functional integral, generally required to make it a mathematically
well defined object, is discussed only briefly. In particular, the
possibility of formulating the argument using a lattice regularization
is mentioned, but not discussed in detail. The general validity of
this statement is called into question by the existence of examples of
spontaneous breaking of vector flavor symmetry on the lattice, namely
in the Aoki phase~\cite{Aoki:1983qi,Aoki:1986xr,Setoodeh:1988ds,
  Aoki:1989rw,Aoki:1990ap,Aoki:1992nb,Aoki:1995yf,Aoki:1995ft,
  Aoki:1995ba,Aoki:1996pw,Aoki:1996af,Aoki:1997fm,Bitar:1996kc,
  Bitar:1997as,Bitar:1997ic,Edwards:1997sp,Edwards:1998sh,
  Sharpe:1998xm,Azcoiti:2008dn,Sharpe:2008ke,Azcoiti:2012ns} of
lattice gauge theories with Wilson fermions~\cite{Wilson1977}. 
While this is not in contradiction with the argument of Vafa and
Witten in the continuum~\cite{Sharpe:1998xm}, it also makes clear that
this argument does not trivially extend to the lattice in a general
setting. It would then be desirable to identify conditions that
guarantee the impossibility of baryon number and vector flavor
symmetry breaking on the lattice, at least for small lattice spacing,
which could help in putting Vafa and Witten's ``theorem'' on more
solid ground.

The strategy of widest generality is to directly prove a lattice
version of Vafa and Witten's bound on the propagator, which would
allow one to recover all the conclusions of Ref.~\cite{Vafa:1983tf} in
a rigorous way (under the tacit assumption of the existence of the
continuum limit). This was done for staggered
fermions~\cite{Kogut:1974ag,Susskind:1976jm,Banks:1976ia} in
Ref.~\cite{Aloisio:2000rb}, so excluding completely the possibility of
breaking baryon number symmetry and the vector flavor symmetry of
several staggered fields on the lattice using this discretization.
However, in four dimensions one flavor of staggered fermions on the
lattice describes four degenerate ``tastes'' of fermions in the
continuum limit, and while the spontaneous breaking of the
corresponding extended flavor symmetry is excluded by the result of
Ref.~\cite{Aloisio:2000rb}, this limits the impossibility proof to a
number of physical fermion species that is a multiple of four (and of
$2^{\left[{d}/{2}\right]}$ in $d$ dimensions). The extension to an
arbitrary number of fermion species requires the ``rooting
trick''~\cite{Hamber:1983kx,Fucito:1984nu,Gottlieb:1988gr} to
eliminate the taste degeneracy, a procedure that has been criticized
in the past (see Refs.~\cite{Creutz:2006ys,Creutz:2006wv,
  Creutz:2007yg,Creutz:2007pr,Creutz:2007rk,Creutz:2008kb}). While
both theoretical arguments and numerical evidence support the validity
of the rooting procedure (see Refs.~\cite{Shamir:2004zc,
  Bernard:2006zw,Bernard:2006vv,Shamir:2006nj,Bernard:2007eh,
  Adams:2008db,Bernard:2008gr}, the reviews~\cite{Durr:2005ax,
  Sharpe:2006re,Bernard:2006qt,Kronfeld:2007ek,Golterman:2008gt}, and
references therein), its theoretical status is still not fully
settled. It would then be nice to extend the proof of
Ref.~\cite{Aloisio:2000rb} or derive a similar bound also for other
discretizations that describe a single fermion species.  However, the
proof makes essential use of the anti-Hermiticity and ultralocality of
the operator: while it can probably be extended quite
straightforwardly to other discretizations that share these
properties, e.g., the minimally doubled fermions of Karsten and
Wilczek~\cite{Karsten:1981gd,Wilczek:1987kw} and of Creutz and Bori{\c
  ci}~\cite{Creutz:2007af,Borici:2007kz} (that are, however, still
describing two fermion species in the continuum limit), it is not
clear how to do so with discretizations that do not, e.g.,
Ginsparg-Wilson fermions~\cite{Ginsparg:1981bj,Hasenfratz:1993sp,
  DeGrand:1995ji,Hasenfratz:1998ri,Kaplan:1992bt,Shamir:1993zy,
  Narayanan:1993sk,Narayanan:1993ss,Neuberger:1997fp,Neuberger:1998wv}.

A less general strategy, still sufficient to prove the impossibility
of spontaneous symmetry breaking on the lattice, is to show that the
corresponding order parameters must vanish. Partial results for vector
flavor symmetry following this strategy are present in the
literature. Already in Ref.~\cite{Vafa:1983tf} the authors show that
vector flavor symmetry cannot be spontaneously broken by the
formation of the simplest symmetry-breaking bilinear fermion
condensate, when approaching the symmetric case of degenerate fermion
masses starting from the non-degenerate case. Their argument works
only for discretizations of the Dirac operator that are
anti-Hermitean, so it applies again only to staggered and minimally
doubled (and obviously to naive) fermions. In
Ref.~\cite{Azcoiti:2010ns} the authors show that the simplest
symmetry-breaking condensate must vanish also for Ginsparg-Wilson
fermions. They do not add any symmetry-breaking term to the action,
applying instead the formalism of probability distribution
functions~\cite{Azcoiti:1995dq,Azcoiti:2008nq} to the relevant
operator to show the absence of degenerate vacua. More precisely,
their result shows that if degenerate non-symmetric vacua are present,
they cannot be distinguished by the (vanishing) expectation value of
this operator.

In this paper I pursue this second strategy and present a simple
argument that spontaneous vector flavor symmetry breaking is
impossible on the lattice for gauge theories with a positive
integration measure, as long as the discretization of the Dirac
operator satisfies certain reasonable assumptions. More precisely, I
show that any localized order parameter for vector flavor symmetry
breaking must vanish in the symmetric limit of fermions of equal
masses (taken of course after the thermodynamic limit), for massive
lattice Dirac operators $\Dirac_M$ that
\begin{enumerate}
   \setcounter{enumi}{-1}
 \item are linear in the fermion masses, $\Dirac_M=\dml + M\deltaD$,
   with $\dml$ and $\deltaD$ trivial in flavor space, and $M$ a
   Hermitean mass matrix;

 \item have inverse bounded in norm by a configuration- and
   volume-in\-de\-pen\-dent constant, finite in the symmetric limit;

 \item have derivative with respect to the fermion masses, $\deltaD$,
   also bounded in norm by a configuration- and
   volume-in\-de\-pen\-dent constant, finite in the symmetric limit.
\end{enumerate}
Assumption (0.)\ is rather natural, and assumption (2.)\ is not really
restrictive; both are satisfied by all common discretizations.
Assumption (1.)\ is instead crucial, and it means that the propagator
corresponding to $\Dirac_M$ is bounded in norm for all configurations,
uniformly in the volume.  This may in general not be the case, for
example if a finite density of near-zero modes of $\Dirac_M$ develops
in the thermodynamic limit, as it happens with Wilson fermions in the
Aoki phase. For staggered\footnote{I refer here to the exact flavor
  symmetry of several staggered fields with the same mass, that holds
  at any lattice spacing, not to the approximate taste symmetry that
  becomes exact only in the continuum limit.}, minimally doubled, and
Ginsparg-Wilson fermions\footnote{The result applies to
  $\gamma_5$-Hermitean Ginsparg-Wilson fermions with $2R=\mathbf{1}$,
  see Section \ref{sec:gwf}.}, assumption (1.)\ holds as long as the
fermion masses are nonzero, and the functional-integration measure is
positive for nonnegative fermion masses, so that for these
discretizations the spontaneous breaking of vector flavor symmetry is
impossible at finite positive fermion mass.

My argument is clearly of narrower scope than the one in
Ref.~\cite{Vafa:1983tf} and its counterpart for staggered fermions in
Ref.~\cite{Aloisio:2000rb}, and limited to quadratic fermion actions
with the usual symmetry-breaking terms. On the other hand, it is
mathematically rigorous for a physicist's standard, leaving little
room for loopholes, and applies to more general discretizations than
staggered fermions. The strategy of proof is standard: one starts from
the explicitly broken case with fermions of different masses, and
shows that observables related by a vector flavor transformation have
the same expectation value in the symmetric limit of equal masses,
taken after the infinite-volume limit.  This is achieved by proving
two rather elementary bounds on the fermion propagator and on its
commutator with the generators of the vector flavor symmetry group,
that hold independently of the lattice size under assumptions
(0.)--(2.).\ This results in the magnitude of the difference between
the expectation values of observables related by a vector flavor
transformation obeying a bound proportional to the spread in mass of
the fermions, uniformly in the volume. In the symmetric limit such
expectation values are then equal, and any order parameter for
symmetry breaking must therefore vanish.  A few remarks are in order.

(i) The geometry of the lattice, the boundary conditions imposed on
the fields, the type of gauge action, the temperature of the system,
and the value of the lattice spacing and of the other parameters of
the theory play no role as long as positivity of the integration
measure and the boundedness assumptions (1.)\ and (2.)\ (or more
generally the derived bounds on the propagator and on its commutator
with the symmetry generators) hold.

(ii) The restriction to localized observables is natural, as
Goldstone's theorem involves observables that are localized in
spacetime, and in space in the finite temperature
case~\cite{Strocchi:2008gsa}. Their counterparts on a finite lattice
involve lattice fields associated with a finite number of lattice
sites or edges (links), that remains unchanged as the system size
grows. In particular, this means that they are polynomial in the
fermion fields, of degree independent of the lattice size.

(iii) If assumptions (0.)--(2.)\ hold for any lattice spacing, or at
least for any sufficiently small spacing, then all the relevant order
parameters vanish in the symmetric infinite-volume theory also in the
continuum limit, if this exists (notice the order of limits:
thermodynamic first, then symmetric, continuum last). Vector flavor
symmetry will then be realized in the continuum. For staggered,
minimally doubled, and Ginsparg-Wilson fermions this is the case for
any positive fermion mass.

(iv) The fate of vector flavor symmetry in the chiral limit, both on
the lattice and in the continuum, can be discussed following the
argument presented in Ref.~\cite{Vafa:1983tf}: barring accidental
degeneracies of the ground states, vector flavor symmetry must remain
unbroken.

(v) The restriction to quadratic actions is not a limitation as far as
the eventual continuum limit is concerned. Renormalizable higher-order
operators with the right global and local symmetries are available
only in dimension lower than or equal to two, where spontaneous
breaking of a continuous symmetry is forbidden~\cite{Mermin:1966fe,
  Hohenberg:1967zz, Coleman:1973ci}. The inclusion of
symmetry-breaking non-renormalizable operators in the action may lead
to spontaneously broken phases on the lattice, but does not affect the
long-distance physics in the continuum limit.  Since lattice
discretizations exist that guarantee the realization of vector flavor
symmetry in the continuum limit, any hypothetical phase where it is
spontaneously broken on the lattice should shrink as this limit is
approached. This is the case also for the spontaneously broken phases
possibly appearing on the lattice for discretizations that do not
satisfy the assumptions of this paper, e.g., the Aoki phase found with
Wilson fermions.

(vi) The existence of regularizations unaffected by residual fermion
doubling in the continuum limit for which the symmetry cannot break
spontaneously on the lattice at any spacing (e.g., Ginsparg-Wilson
fermions) establishes rigorously (at the physicist's level of rigor)
the impossibility of its spontaneous breaking in continuum gauge
theories for any number of physical fermion species.

The plan of the paper is the following. After briefly reviewing gauge
theories on the lattice to set up the notation in Section
\ref{sec:not}, and vector flavor symmetry in Section
\ref{sec:vftransf}, I derive the relevant bounds and prove the main
statement in Section \ref{sec:bounds}. The cases of staggered,
Ginsparg-Wilson, Wilson, and minimally doubled fermions are discussed
in Section~\ref{sec:examples}. A brief summary is given in
Section~\ref{sec:concl}. A few technical details are given in
Appendix~\ref{sec:app}.

\section{Gauge theories on the lattice}
\label{sec:not}

I will consider $d$-dimensional vector gauge theories with $N_f$
flavors of fermions, all transforming in the same $N_c$-dimensional
representation of a compact gauge group, discretized on a finite
lattice containing $\Vc$ sites.  Suitable boundary conditions are
assumed on the gauge and fermion fields. The shape of the lattice and
the boundary conditions play no distinctive role in the following; in
particular, the discussion applies to systems both at zero and finite
temperature. The partition function and the expectation values of the
theory are given by
\begin{equation}
  \label{eq:partfunc_expval}
  \begin{aligned}
    Z &\equiv \int [\gmeas]\int [\fmeas]\,
    e^{-\act_{\mathrm{G}}[U] - \act_{\mathrm{F}}[\psi,\bar{\psi},U]}\,,\\
    \la \Oc \ra &\equiv \f{1}{Z} \int [\gmeas]\int [\fmeas]\,
    e^{-\act_{\mathrm{G}}[U] -
      \act_{\mathrm{F}}[\psi,\bar{\psi},U]}\Oc[\psi,\bar{\psi},U]\,,\\
  \end{aligned}
\end{equation}
where $[\gmeas]=\prod_{\ell} dU_\ell$ is the product of the Haar
measures associated with the gauge variables $U_\ell$ attached to the
lattice links $\ell$, and
$[\fmeas]=\prod_{x f a\alpha} d\psi_{f a\alpha}(x) d\bar{\psi}_{f
  a\alpha}(x)$ is the Berezin integration measure associated with the
Grassmann variables $\psi_{f a\alpha}(x)$ and
$\bar{\psi}_{f a\alpha}(x)$ attached to the lattice sites $x$.  Here
$f$ and $a$ are the discrete indices associated with the flavor and
color (i.e., gauge group) degrees of freedom, $f=1,\ldots, N_f$,
$a=1,\ldots,N_c$, and $\alpha$ is the Dirac index, typically
$\alpha=1,\ldots,2^{\left[{d}/{2}\right]}$, but possibly absent
altogether (e.g., for staggered fermions). The full set of discrete
indices will be collectively denoted as $A=f a\alpha$; when needed,
the color and Dirac indices will be denoted together as
$\Asc=a\alpha$. Finally, $\act_{\mathrm{G}}$ and $\act_{\mathrm{F}}$
denote the gauge and fermionic parts of the action. The fermionic
action is taken to be of the form
\begin{equation}
  \label{eq:ferm_act}
  \begin{aligned}
    \act_{\mathrm{F}}[\psi,\bar{\psi},U] &=
    \sum_{x,y,A,B}\bar{\psi}_A(x) \left(\Dirac_M[U]\right)_{AB}(x,y)
    \psi_B(y) \\ &=\bar{\psi} \Dirac_M[U] \psi\,,
  \end{aligned}
\end{equation}
where in the last passage I introduced the matrix notation that will
be used repeatedly. Here $\Dirac_M$ is the massive Dirac operator,
whose dependence on the gauge links $U_\ell$ will be often omitted for
simplicity. Expectation values are computed in two steps. For a
generic observable $\Oc[\psi,\bar{\psi},U]$, integration over
Grassmann variables yields
\begin{equation}
  \label{eq:gen_obs_F_generic}
  \begin{aligned}
    \left\la \Oc\right\ra_{\mathrm{F}} &\equiv \f{\int [\fmeas]\,
      e^{-\act_{\mathrm{F}}[\psi,\bar{\psi},U]}
      \Oc[\psi,\bar{\psi},U]}{\int [\fmeas]\, e^{-
        \act_{\mathrm{F}}[\psi,\bar{\psi},U]} } \\ &= \f{\int
      [\fmeas]\, e^{-\act_{\mathrm{F}}[\psi,\bar{\psi},U]}
      \Oc[\psi,\bar{\psi},U]}{\det\Dirac_M[U] } \,.
  \end{aligned}
\end{equation}
The expectation value of $\Oc$ is then obtained from
Eq.~\eqref{eq:gen_obs_F_generic} by averaging over gauge fields,
$ \la \Oc\ra = \la \la\Oc\ra_{\mathrm{F}}\ra_{\mathrm{G}}$, where for
a purely gluonic observable $\tilde{\Oc}[U]$
\begin{equation}
  \label{eq:gen_obs_G}
  \begin{aligned}
    \la \tilde{\Oc} \ra_{\mathrm{G}} &\equiv \f{\int
      [\gmeas]\,e^{-\act_{\mathrm{G}}[U]}\det\Dirac_M[U]
      \tilde{\Oc}[U]} {\int
      [\gmeas]\,e^{-\act_{\mathrm{G}}[U]}\det\Dirac_M[U]} \\ &
    = \f{1}{Z}
     \int [\gmeas]\,e^{-\act_{\mathrm{G}}[U]}\det\Dirac_M[U]
    \tilde{\Oc}[U] \,.
  \end{aligned}
\end{equation}
I assume that the full gluonic integration measure
$d\mu_{\mathrm{G}}=[\gmeas]\,e^{-\act_{\mathrm{G}}[U]}\det\Dirac_M[U]$
is nonnegative, i.e.,
$e^{-\act_{\mathrm{G}}[U]}\det\Dirac_M[U]\ge 0$, and not identically
zero. For brevity, I will refer to this assumption simply as
positivity of the integration measure. The gluonic action is
otherwise unspecified, besides its being gauge-invariant. I consider
massive Dirac operators of the form
\begin{equation}
  \label{eq:dirop}
    \Dirac_M = \mathbf{1}_F \dml + M\deltaD\,,
\end{equation}
with $M$ a constant Hermitean matrix carrying only flavor indices and
independent of coordinates and gauge links. The symbol $\mathbf{1}_F$,
and similarly $\mathbf{1}_{C}$ and $\mathbf{1}_{D}$, denote the
identity in flavor ($F$), color ($C$), and Dirac ($D$) space;
$\mathbf{1}$ will denote the identity in the full flavor, color, Dirac
and coordinate space. The operators $\dml$ and $\deltaD$ carry only
color, Dirac, and coordinate indices, i.e., $(\dml)_{\Asc \Bsc}(x,y)$
and $(\deltaD)_{\Asc \Bsc}(x,y)$. Since one can diagonalize $M$ with a
unitary transformation, and reabsorb this into a redefinition of the
fermion fields that does not affect the Berezin integration measure,
one can consider a diagonal mass matrix $M=\diag(m_1,\ldots,m_{N_f})$
without loss of generality, and write
\begin{equation}
  \label{eq:dirop2}
  \begin{aligned}
    \Dirac_M &=
    \diag\left(D^{(m_1)},\ldots,D^{(m_{N_f})}\right)\,, \\
    D^{(m)} &\equiv \dml + m\deltaD\,,\\
    (\Dirac_M)_{f\Asc\,g\Bsc}(x,y) &= \delta_{fg}
    (D^{(m_f)})_{\Asc\Bsc}(x,y)\,.
  \end{aligned}
\end{equation}
The fermion propagator is then
\begin{equation}
  \label{eq:propo0}
  \begin{aligned}
    \propa_M&=\Dirac_M^{-1}
    =\diag\left(S^{(m_1)},\ldots, S^{(m_{N_f})}\right) \,, \\
    S^{(m)}&\equiv (D^{(m)})^{-1} \,,
    \\
    (\propa_M)_{f\Asc \,g\Bsc}(x,y) &= \delta_{fg} S^{(m_f)}_{\Asc
      \Bsc}(x,y)\,,
  \end{aligned}
\end{equation}
and the fermion determinant is
\begin{equation}
  \label{eq:det_nf}
  \det \Dirac_M = {\prod_{f=1}^{N_f}}\det D^{(m_f)}\,.
\end{equation}
The trace over all indices, i.e., flavor, color, Dirac, and
coordinates, will be denoted by $\Tr$. The trace over one or more of
the discrete indices will be denoted by $\tr$ with one or more of the
subscripts $F$, $C$, $D$, indicating which indices are being traced.
Matrix multiplication is understood not to involve the indices
displayed explicitly: for matrices $\mathrm{P},\mathrm{Q}$ carrying
all indices, and matrices $P,Q$ carrying all but flavor indices,
\begin{equation}
  \label{eq:matmultexamples}
  \begin{aligned}
    \left(\mathrm{P}\mathrm{Q}\right)_{AB}(x,y)&= \sum_{z,C}
    \mathrm{P}_{AC}(x,z)\mathrm{Q}_{CB}(z,y)\,,\\
    \left(\mathrm{P}(x,z)\mathrm{Q}(z,y)\right)_{AB}&= \sum_{C}
    \mathrm{P}_{AC}(x,z)\mathrm{Q}_{CB}(z,y)\,,\\
    \left(PQ\right)_{\Asc\Bsc}(x,y)&= \sum_{\Cscc,z}
    P_{\Asc\Cscc}(x,z)Q_{\Cscc\Bsc}(z,y)    \,,\\
    \left(P(x,z)Q(z,y)\right)_{\Asc\Bsc}&= \sum_{\Cscc}
    P_{\Asc\Cscc}(x,z)Q_{\Cscc\Bsc}(z,y) \,.
  \end{aligned}
\end{equation}
A similar convention applies to Hermitean conjugation,
\begin{equation}
  \label{eq:hermconjgexamples}
  \begin{aligned}
    (\mathrm{P}^\dag)_{AB}(x,y) &=
    \mathrm{P}_{BA}(y,x)^*    \,, \\
    (\mathrm{P}(x,y)^\dag)_{AB} &=
    \mathrm{P}_{BA}(x,y)^*\,,\\
    (P^\dag)_{\Asc\Bsc}(x,y) &= P_{\Bsc\Asc}(y,x)^*\,,\\
    (P(x,y)^\dag)_{\Asc\Bsc} &= P_{\Bsc\Asc}(x,y)^*\,.
  \end{aligned}
\end{equation}
At a certain point I will assume that the propagator and the operator
$\deltaD$ are suitably bounded in norm. In the finite-dimensional
case, the operator norm $\Vert \mathrm{A}\Vert$ of an operator
$\mathrm{A}$ equals the largest of the eigenvalues $a_n^2$ of the
positive Hermitean operator $\mathrm{A}^\dag \mathrm{A}$,
\begin{equation}
  \label{eq:opnorm0}
  \begin{aligned}
    \Vert \mathrm{A} \Vert^2 &= \sup_{\psi\neq
      0}\f{(\mathrm{A}\psi,\mathrm{A}\psi)}{(\psi,\psi)} =
    \sup_{\psi\neq 0}\f{(\psi,\mathrm{A}^\dag
      \mathrm{A}\psi)}{(\psi,\psi)} 
    = \max_{n} a_n^2\,,
  \end{aligned}
\end{equation}
where $(\psi,\phi)$ denotes the standard Hermitean inner product. I
will assume that
\begin{enumerate}
\item $\Vert \propa_M\Vert\le m_0^{-1}< \infty$, with $m_0$
  independent of the gauge configuration and of the lattice volume
  $\Vc$, and finite in the limit of equal fermion masses;
\item $\Vert \deltaD\Vert \le \ded<\infty$, with $\ded$ independent of
  the gauge configuration and of the lattice volume, and finite in the
  limit of equal fermion masses.
\end{enumerate}

For the purposes of this paper it suffices to consider the most
general localized gauge-invariant observable, so polynomial in the
fermion fields and dependent on finitely many link variables, with
fermion number zero\footnote{On a finite lattice observables are
  necessarily polynomial in the fermion fields and dependent on
  finitely many link variables: the restriction is relevant in the
  thermodynamic limit. Nonzero fermion number immediately entails a
  vanishing expectation value. Any observable that is not
  gauge-invariant can be replaced with its average over gauge orbits
  without changing its expectation value.}.  For notational purposes
it is convenient to write it with its discrete indices contracted with
the most general matrix carrying flavor, color and Dirac indices,
dependent on the link variables (and possibly also explicitly on the
lattice coordinates, and on the parameters of the theory), and having
the right transformation properties under gauge transformations to
make the observable gauge invariant. I will then consider
\begin{widetext}
  \begin{equation}
  \label{eq:gen_obs_definition}
\begin{aligned}
    \Oc_{\Mc}[\psi,\bar{\psi},U] 
  &\equiv {\sum_{\substack{A_1,\ldots,A_n,\\B_1,\ldots,B_n}}}
  (\Mc[U])_{A_1\ldots A_n B_1
    \ldots B_n}(x_1,\ldots,x_n,y_1,\ldots,y_n) 
{\prod_{i=1}^n}
  \psi_{B_i}(y_i) \bar{\psi}_{A_i}(x_i) \\ & =
  \sum_{\mathbf{A},\mathbf{B}}(\Mc[U])_{\mathbf{A}\mathbf{B}}(\mathbf{x},\mathbf{y})
  { \prod_{i=1}^n} \psi_{B_i}(y_i) \bar{\psi}_{A_i}(x_i) \,.
\end{aligned}
\end{equation}
\end{widetext}
For brevity I will write $\Mc[U]_{\colA\colB}(\colx,\coly)$, using
bold typeface to denote collectively a set of indices or variables. I
will generally omit the dependence on $U$ when unimportant. The
transformation properties of $\Mc$ under gauge transformations are
easily obtained from those of the fermionic fields, and do not play
any role in the following. Notably, the quantity
\begin{equation}
  \label{eq:K_def}
  \Kc_\Mc(\colx,\coly) \equiv \tr_{FCD}\{\Mc(\colx,\coly)
  \Mc(\colx,\coly)^\dag\}
\end{equation}
is gauge invariant. The expectation value
$ \la \Oc_\Mc\ra = \la \la\Oc_\Mc\ra_{\mathrm{F}}\ra_{\mathrm{G}}$ is
obtained averaging $\la\Oc_\Mc\ra_{\mathrm{F}}$ over gauge fields
using Eq.~\eqref{eq:gen_obs_G}. From Eq.~\eqref{eq:gen_obs_F_generic}
one finds using Wick's theorem
  \begin{equation}
  \label{eq:gen_obs_F}
  \begin{aligned}
    \left\la \Oc_{\Mc}\right\ra_{\mathrm{F}} & =
    \sum_{\mathbf{A},\mathbf{B}}\Mc[U]_{\colA\colB}(\colx,\coly)
    \left\la \prod_{i=1}^n \psi_{B_i}(y_i)\bar{\psi}_{A_i}(x_i)
    \right\ra_{\mathrm{F}} \\ & =
    \sum_{\mathbf{A},\mathbf{B}}\Mc[U]_{\colA\colB}(\colx,\coly) \\
    &\phantom{=}\times \sum_{\mathrm{P}\in
      \mathrm{S}_n}\sigma_{\mathrm{P}} \prod_{i=1}^n \propa_M[U]_{B_i
      A_{\mathrm{P}(i)}}(y_i,x_{\mathrm{P}(i)})\,,
  \end{aligned}
\end{equation}
with $\mathrm{P}$ a permutation of $n$ elements and
$\sigma_{\mathrm{P}}=\pm 1$ its signature.

Restrictions on $\Mc$ are required in order for the integration over
link variables to yield finite results for $\la \Oc_\Mc\ra$. In the
physically relevant cases $\Mc$ is a product of Wilson lines, suitably
connecting the fermion fields to achieve gauge invariance, and so
polynomial in the link variables. Imposing that $\Mc$ be polynomial
or, more generally, continuous in the link variables guarantees that
in a finite volume $\Kc_\Mc$ is bounded from above by its maximum on
the compact integration manifold. The thermodynamic limit is taken
while keeping $\Mc$ fixed as a function of the link variables (in
particular, its possible dependence on the lattice coordinates is
unchanged and cannot cause convergence problems), and so the bound on
$\Kc_\Mc$ is independent of the volume.  Assumption (1.)\ on the
propagator then suffices to show convergence of $\la \Oc_\Mc\ra$, both
in a finite volume and in the infinite-volume limit, see below in
Section \ref{sec:bounds}. Finally, any possible dependence of $\Mc$ on
the fermion masses is assumed to be continuous, at least in the
symmetric limit of equal fermion masses. This guarantees that
$\Kc_\Mc$ is bounded in a neighborhood of the symmetric point, which
is all that is needed to prove the results of Section
\ref{sec:bounds}. In fact, the assumption of continuity of $\Mc$ in
the link variables and in the fermion masses can be relaxed, without
changing the arguments in Section \ref{sec:bounds}, to the weaker
assumption that $\Kc_\Mc$ be bounded from above, independently of the
link configuration (and therefore of the volume), in a neighborhood of
the symmetric point. The results of this paper can be proved also if
one further relaxes the requirement of continuity or boundedness to
absolute integrability of the entries $\Mc_{\colA\colB}$: this is
discussed in Appendix \ref{sec:app}.

\section{Vector flavor transformations}
\label{sec:vftransf}

Vector flavor transformations are defined by
\begin{equation}
  \label{eq:wi0}
  \begin{aligned}
    \psi_{f\Asc}(x)&\to
    \textstyle{  \sum_g}V_{fg}\psi_{g\Asc}(x), \\
    \bar{\psi}_{f\Asc}(x)&\to \textstyle{
      \sum_g}\bar{\psi}_{g\Asc}(x)V^\dag_{gf}\,,
  \end{aligned}
\end{equation}
where $V_{fg}$ are the entries of a unitary unimodular $N_f\times N_f$
matrix $V \in \mathrm{SU}(N_f)$.  This can be written as
$V=e^{i\theta_a t^a}\equiv e^{i\bm{\theta}\cdot\bm{t}}\equiv
V(\bm{\theta})$, with $\theta_a\in\mathbb{R}$ and with $t^a$ the
Hermitean and traceless generators of $\mathrm{SU}(N_f)$, taken with
the standard normalization $2\,\tr_F\, t^at^b = \delta^{ab}$. The task
is to show that any localized observable $\Oc=\Oc[\psi,\bar{\psi},U]$
and its transformed $\Oc^{\bm{\theta}}$,
\begin{equation}
  \label{eq:w1_not}
  \Oc^{\bm{\theta}}[\psi,\bar{\psi},U]\equiv
  \Oc[V(\bm{\theta})\psi,\bar{\psi}V(\bm{\theta})^\dag,U]\,, 
\end{equation}
have the same expectation value in the infinite-volume theory in the
symmetric limit $M\to m\mathbf{1}_F$, i.e.,
\begin{equation}
  \label{eq:task}
  \lim_{M\to m\mathbf{1}_F}\lim_{\Vc\to\infty}\left\la
    \Oc^{\bm{\theta}}-\Oc\right\ra 
  = 0\,.
\end{equation}
On a finite lattice all observables are obviously localized. In the
thermodynamic limit $\Vc\to\infty$, every localized observable is a
linear combination of finitely many of the $\Oc_\Mc$ discussed above,
Eq.~\eqref{eq:gen_obs_definition}, where $\Mc$ is understood to be a
fixed function of the link variables, independent of $\Vc$. Moreover,
since $\mathrm{SU}(N_f)$ is a Lie group, any finite transformation can
be obtained by composition of infinitesimal ones. It suffices then to
consider observables $\Oc_\Mc$ and transformations with
$\theta_a\ll 1$ in Eq.~\eqref{eq:task}, i.e., one has to show
\begin{equation}
  \label{eq:task_inf}
  \lim_{M\to m\mathbf{1}_F}\lim_{\Vc\to\infty}\f{\de}{\de
    \theta_a}\left\la\Oc_\Mc^{\bm{\theta}}\right\ra \bigg|_{\bm{\theta}=0}
  = 0\,.
\end{equation}
An explicit proof that Eq.~\eqref{eq:task_inf} implies
Eq.~\eqref{eq:task} is given in Appendix \ref{sec:finflavtr}. In
Appendix \ref{sec:orpar} I show that Eq.~\eqref{eq:task} implies that
order parameters for vector flavor symmetry, i.e., expectation values
of observables that transform nontrivially under $\mathrm{SU}(N_f)$,
must vanish in the symmetric limit.

To efficiently study the effect of a vector flavor transformation on
the expectation value of an arbitrary observable, it is convenient to
make use of the corresponding well-known integrated Ward-Takahashi
identity, derived for completeness in Appendix \ref{sec:finflavtr},
\begin{equation}
  \label{eq:wi3}
  \begin{aligned}
    i\f{\de}{\de\theta_a}\left\la \Oc^{\bm{\theta}}\right\ra
    \bigg|_{\bm{\theta}=0}&= \left\la \Cc^a \Oc\right\ra\,, \\
    \Cc^a[\psi,\bar{\psi},U] &\equiv \bar{\psi}[ t^a,
    \Dirac_M[U]]\psi\,.
  \end{aligned}
\end{equation}
Integrating fermions out one finds for a generic observable $\Oc$
\begin{equation}
  \label{eq:wi5}
  \begin{aligned}
    i\f{\de}{\de\theta_a}\left\la
      \Oc^{\bm{\theta}}\right\ra_{\mathrm{F}} \bigg|_{\bm{\theta}=0}
    &= \left\la \Cc^a \Oc\right\ra_{\mathrm{F}} = \left\la
      \Cc^a\ra_{\mathrm{F}} \la \Oc\right\ra_{\mathrm{F}} - \Tr\left\{
      F^a O\right\} \,,
      \end{aligned}
\end{equation}
with
\begin{equation}
  \label{eq:def_Oder}
  O_{A B}(x,y)\equiv
  \left\la\f{\de}{\de_L\psi_{B}(y)}\f{\de}{\de_L\bar{\psi}_{A}(x)}
    \Oc\right\ra_{\mathrm{F}} \,,
\end{equation}
where $\de_{L}$ denotes the usual left derivative with respect to
Grassmann variables, and
\begin{equation}
  \label{eq:commF}
  \begin{aligned}
    F^a\equiv \propa_M [t^a, \Dirac_M] \propa_M= [\propa_M,t^a]\,,
  \end{aligned}
\end{equation}
i.e., the commutator of the propagator with the generators of
$\mathrm{SU}(N_f)$. The first term in Eq.~\eqref{eq:wi5} vanishes
since 
\begin{equation}
  \label{eq:wi5_firstterm}
  \begin{aligned}
    \left\la \Cc^a\right\ra_{\mathrm{F}} & = - \Tr\left\{[ t^a,
      \Dirac_M] \propa_M\right\} = - \Tr\left\{ t^a \left[\Dirac_M,
        \propa_M\right]\right\} =0\,.
  \end{aligned}
\end{equation}
Specializing now to observables of the form
Eq.~\eqref{eq:gen_obs_definition}, one finds, for $i,j=1,\ldots,n$,
\begin{equation}
  \label{eq:wi11bis_reorder}
  \begin{aligned}
    & \f{\de}{\de_L\psi_{B_j}(y_j)}\f{\de}{\de_L\bar{\psi}_{A_i}(x_i)}
    \Oc_\Mc[\psi,\bar{\psi},U] \\ & = -s_{ij}
    (\Mc[U])_{\mathbf{A}\mathbf{B}}(\mathbf{x},\mathbf{y})
    \prod_{k=1}^{n-1} \psi_{B_k^{(j)}}(y_k^{(j)})
    \bar{\psi}_{A_k^{(i)}}(x_k^{(i)}) \,.
  \end{aligned}
\end{equation}
Here the superscript $(i)$ means that the $i$th element is omitted
from the set of indices while keeping their ordering unchanged, i.e.,
$ \colA^{(i)}= \{{A}^{(i)}_1,\ldots,{A}^{(i)}_{n-1}\} =
\{A_1,\ldots,A_{i-1}\}\cup \{A_{i+1},\ldots,A_{n}\} $, and similarly
for the other sets.  The sign factor $-s_{ij}=(-1)^{i-j-1}$ appears
when reordering the Grassmann variables to be in the same form as in
Eq.~\eqref{eq:gen_obs_definition}. Using now Eq.~\eqref{eq:gen_obs_F}
one finds explicitly
\begin{widetext}
\begin{equation}
  \label{eq:wi11bis}
  \begin{aligned}
    O_{\Mc\,A_i B_j}(\colx,\coly)&\equiv \left\la
      \f{\de}{\de_L\psi_{B_j}(y_j)}\f{\de}{\de_L\bar{\psi}_{A_i}(x_i)}
      \Oc_\Mc \right\ra_{\mathrm{F}} = -s_{ij} \sum_{\mathrm{P}\in
      \mathrm{S}_{n-1}}\sigma_{\mathrm{P}}\sum_{{\colA}^{(i)},{\colB}^{(j)}}
    \Mc_{\colA\colB}(\colx,\coly) \prod_{k=1}^{n-1}
    (\propa_M)_{{B}^{(j)}_k{A}^{(i)}_{\mathrm{P}(k)}}({y}^{(j)}_k,{x}^{(i)}_{\mathrm{P}(k)})
    \,,
  \end{aligned}
\end{equation}
\end{widetext}
where $\mathrm{P}$ is now a permutation of $n-1$ elements and
$\sigma_{\mathrm{P}}$ its signature. Notice that $A_i$ and $B_j$ are
not contracted in Eq.~\eqref{eq:wi11bis}, and that the dependence of
$ O_{\Mc\,A_i B_j}$ on $x_i,y_j$ is only through that of $\Mc$.  One
can finally write
\begin{widetext}
\begin{equation}
  \label{eq:wi11quater}
  \begin{aligned}
    & i\f{\de}{\de\theta_a}\left\la \Oc_\Mc^{\bm{\theta}}\right\ra
    \Big|_{\bm{\theta}=0}= \left\la \Cc^a \Oc_\Mc\right\ra =
    \sum_{i,j=1}^n s_{ij} \sum_{\mathrm{P}\in
      \mathrm{S}_{n-1}}\sigma_{\mathrm{P}}\left\la
      \sum_{\colA,\colB}\Mc_{\colA\colB}(\colx,\coly) F^a_{B_j
        A_i}(y_j,x_i)\prod_{k=1}^{n-1}
      (\propa_M)_{{B}^{(j)}_k{A}^{(i)}_{\mathrm{P}(k)}}({y}^{(j)}_k,{x}^{(i)}_{\mathrm{P}(k)})
    \right\ra_{\mathrm{G}}\,.
\end{aligned}
\end{equation}
\end{widetext}
This relation could have been obtained also directly from
Eq.~\eqref{eq:gen_obs_F} by noticing that a vector flavor
transformation $V(\bm{\theta})$ can be seen as a transformation of the
matrix $\Mc$, $\Mc\to\Mc^{\bm{\theta}}$, at fixed fermion fields,
i.e., $\Oc_\Mc^{\bm{\theta}}=\Oc_{\Mc^{\bm{\theta}}}$ [see
Eq.~\eqref{eq:wi1bis_app2}].  In practice,
$\la \Oc_{\Mc^{\bm{\theta}}}\ra_{\mathrm{F}}$ is obtained from
$\la\Oc_{\Mc}\ra_{\mathrm{F}}$ by replacing each propagator with
$\propa_M\to V(\bm{\theta})\propa_M V(\bm{\theta})^\dag$. Expanding to
first order in $\bm{\theta}$, Eq.~\eqref{eq:wi11quater} follows.

\section{Proof of the main result}
\label{sec:bounds}

Equation~\eqref{eq:wi11quater} is the starting point for establishing
a bound on the magnitude of the variation of the expectation value of
$\Oc_\Mc$ under an infinitesimal vector flavor transformation. The
argument is quite elementary, and should be understandable in spite of
my best attempts at obscuring it with cumbersome notation. As shown
above, after performing the Wick contractions the variation of
$\la\Oc_\Mc\ra$ is written as a sum of terms of the form
$\la \Mc F \prod \propa\ra_{\mathrm{G}}$, differing by permutations of
the indices. By standard techniques, if the integration measure is
positive this is bounded in magnitude by a sum of terms of the form
$\la(\tr\{ \Mc\Mc^\dag\} \tr\{ F F^\dag\} \prod
\tr\{\propa\propa^\dag\})^{\f{1}{2}}\ra_{\mathrm{G}}$, which under the
assumptions of Section \ref{sec:not} on the Dirac operator and on the
continuity (or boundedness) of $\Mc$ can be bounded uniformly in the volume by a
constant times the spread in mass of the fermions. In the symmetric
limit one then finds that the variation of $\la\Oc_\Mc\ra$ under any
infinitesimal vector flavor transformation vanishes.  By a similar
argument one can show that $\la\Oc_\Mc\ra$ is finite, independently of
the fermion masses, also in the thermodynamic limit.

I now present the detailed proof. Using standard inequalities for the
absolute value, the assumed positivity of the integration measure, and
the Cauchy-Schwarz inequality for inner products, one has
\begin{widetext}
\begin{equation}
  \label{eq:cs1}
  \begin{aligned}
    \left| \f{\de}{\de\theta_a}\left\la \Oc_\Mc^{\bm{\theta}}
      \right\ra \Big|_{\bm{\theta}=0}\right| 
    & \le \sum_{i,j=1}^n\sum_{\mathrm{P}\in \mathrm{S}_{n-1}}
    \left|\left\la \sum_{\colA,\colB}\Mc_{\colA\colB}(\colx,\coly)
        F^a_{B_j A_i}(y_j,x_i) \prod_{k=1}^{n-1}
        (\propa_M)_{{B}^{(j)}_k{A}^{(i)}_{\mathrm{P}(k)}}({y}^{(j)}_k,{x}^{(i)}_{\mathrm{P}(k)})
      \right\ra_{\mathrm{G}}\right| \\
    &\le \sum_{i,j=1}^n\sum_{\mathrm{P}\in \mathrm{S}_{n-1}} \left\la
      \left| \sum_{\colA,\colB}\Mc_{\colA\colB}(\colx,\coly) F^a_{B_j
          A_i}(y_j,x_i) \prod_{k=1}^{n-1}
        (\propa_M)_{{B}^{(j)}_k{A}^{(i)}_{\mathrm{P}(k)}}({y}^{(j)}_k,{x}^{(i)}_{\mathrm{P}(k)})
      \right|\right\ra_{\mathrm{G}}  \\
    &\le \sum_{i,j=1}^n\sum_{\mathrm{P}\in S_{n-1}} \left\la\left(
        \Kc_\Mc(\colx,\coly) \Fc^a(y_j,x_i) \prod_{k=1}^{n-1}
        \Sc({y}^{(j)}_k,{x}^{(i)}_{\mathrm{P}(k)})
      \right)^{\f{1}{2}}\right\ra_{\mathrm{G}} \,,
  \end{aligned}
\end{equation}
\end{widetext}
where $\Kc_\Mc(\colx,\coly)$ is defined in Eq.~\eqref{eq:K_def} and
\begin{equation}
  \label{eq:FS_def}
  \begin{aligned}
    \Fc^a(y,x) &\equiv \tr_{FCD}\{ F^a(y,x) F^a(y,x)^\dag\}\,, \\
    \Sc(y,x) &\equiv \tr_{FCD}\{\propa_M(y,x)\propa_M(y,x)^\dag\}\,.
  \end{aligned}
\end{equation}
The quantity $\Kc_\Mc(\colx,\coly)$ is a positive gauge-invariant
function of the link variables and their Hermitean conjugates,
polynomial (or more generally continuous) if $\Mc$ is polynomial
(continuous), defined on the compact domain given by the direct
product of finitely many compact gauge-group manifolds. It is
therefore bounded from above in magnitude by its maximum, which
depends on the details of $\Mc$ but is otherwise a configuration- and
volume-independent quantity,
\begin{equation}
  \label{eq:bound_coeff}
  \Kc_\Mc(\colx,\coly)= \tr_{FCD}
  \{\Mc(\colx,\coly)\Mc(\colx,\coly)^\dag \}\le \Csc_\Mc 
  \,.  
\end{equation}
One can actually relax the request of continuity of $\Mc$ to assuming
that this bound on $\Kc_\Mc$ holds, without changing the argument
below. Moreover, since $\dml$ and $\deltaD$ are trivial and $M$ is
diagonal in flavor space, one has $[t^a,\Dirac_M] = [t^a,M]\deltaD$,
\begin{equation}
  \label{eq:bound_comm_good0}
  \begin{aligned}
    & F^a_{g\Bsc\,f\Asc }(y,x) = (\propa_M [t^a,M] \deltaD
    \propa_M)_{g\Bsc\,f\Asc}(y,x) \\ & =
    t^a_{gf}(m_f-m_g)(S^{(m_g)}\deltaD S^{(m_f)})_{\Bsc\Asc}(y,x)\,,
  \end{aligned}
\end{equation}
and so
\begin{equation}
  \label{eq:bound_comm_good}
  \begin{aligned}
    \Fc^a(y,x) &= \textstyle{\sum_{f,g}} |t^a_{gf}|^2 (m_f-m_g)^2 \\
    &\phantom{= \textstyle{\sum_{f,g}} }\times\tr_{CD}\{
    (S^{(m_g)}\deltaD S^{(m_f)})(y,x) \\ &\phantom{=
      \textstyle{\sum_{f,g}} \tr_{CD}}\times(S^{(m_g)}\deltaD
    S^{(m_f)})(y,x)^\dag \}\,.
  \end{aligned}
\end{equation}
The partial traces appearing in Eqs.~\eqref{eq:FS_def} and
\eqref{eq:bound_comm_good} can be bounded using an elementary lemma,
proved in Appendix \ref{sec:boundpt}: given a multi-indexed matrix
$\mathrm{A}$, the partial trace of $\mathrm{A}^\dag \mathrm{A}$ over a
subset of its indices is bounded by the dimension of the corresponding
space times the square of the operator norm of $\mathrm{A}$. One has
then the following bound on the propagator [see
Eq.~\eqref{eq:lemma2}], valid for arbitrary lattice coordinates $x$
and $y$,
\begin{equation}
  \label{eq:bound_prop_good3}
  \Sc(y,x)   \le  N_f N_c N_D  \Vert \propa_M \Vert^2  \,,
\end{equation}
from which one obtains the bound
\begin{equation}
  \label{eq:bound_prop_good4}
  \begin{aligned}
    \prod_{k=1}^{n-1} \Sc({y}^{(j)}_k,{x}^{(i)}_{\mathrm{P}(k)}) \le
    \left( N_f N_c N_D \Vert \propa_M \Vert^2\right)^{n-1}\,.
  \end{aligned}
\end{equation}
Since the bound Eq.~\eqref{eq:bound_prop_good3} is independent of the
coordinates, the bound Eq.~\eqref{eq:bound_prop_good4} is independent
of the particular choice of $i$ and $j$ and of the permutation
$\mathrm{P}$.  Using again Eq.~\eqref{eq:lemma2} one finds
\begin{equation}
  \label{eq:bound_comm_good2}
  \begin{aligned}
    & \tr_{CD} \{(S^{(m_g)}\deltaD S^{(m_f)})(y,x) (S^{(m_g)}\deltaD
    S^{(m_f)})(y,x)^\dag\} \\ & \le N_c N_D \Vert S^{(m_g)}\deltaD
    S^{(m_f)}\Vert^2 \\ & \le N_c N_D \Vert S^{(m_g)} \Vert^2 \Vert
    \deltaD\Vert^2 \Vert S^{(m_f)}\Vert^2\\ & \le N_c N_D \Vert
    \propa_M \Vert^4 \Vert \deltaD\Vert^2 \,,
  \end{aligned}
\end{equation}
where I used the well-known inequality
$\Vert AB\Vert \le \Vert A\Vert \Vert B\Vert$, and the obvious fact
that
$\Vert S^{(m_f)} \Vert \le \max_{f'} \Vert S^{(m_{f'})} \Vert =\Vert
\propa_M \Vert$.  From Eqs.~\eqref{eq:bound_comm_good} and
\eqref{eq:bound_comm_good2} one obtains the following bound on the
commutator of the propagator with the $\mathrm{SU}(N_f)$ generators,
\begin{equation}
  \label{eq:bound_comm_good3}
  \begin{aligned}
    \Fc^a(y,x) &\le \sum_{f,g} |t^a_{gf}|^2(m_f-m_g)^2 N_c N_D\Vert
    \propa_M \Vert^4 \Vert \deltaD\Vert^2 \\ &\le \tf{1}{2} (\delta
    m)^2 N_c N_D \Vert \propa_M \Vert^4 \Vert \deltaD\Vert^2 \,,
  \end{aligned}
\end{equation}
where I denoted with $\delta m \equiv \max_{f,g}|m_f-m_g|$ the spread
in mass of the fermions, and I used $\tr_F (t^a)^2=\f{1}{2}$.  Also
this bound is independent of the coordinates, and so each of the
$n^2 (n-1)!$ terms appearing in the sums over $i$, $j$ and
$\mathrm{P}$ in Eq.~\eqref{eq:cs1} obeys the same bound.  Collecting
now Eqs.~\eqref{eq:bound_coeff}, \eqref{eq:bound_prop_good4}, and
\eqref{eq:bound_comm_good3}, one finds
\begin{equation}
  \label{eq:bound_summary}
  \begin{aligned}
    \left| \f{\de}{\de\theta_a}\left\la \Oc_\Mc^{\bm{\theta}}
      \right\ra \Big|_{\bm{\theta}=0}\right| &\le \delta m\,
    \tilde{\Csc}_\Mc \left\la \Vert \propa_M \Vert^{n+1} \Vert
      \deltaD\Vert\right\ra_{\mathrm{G}}\,,
  \end{aligned}
\end{equation}
having set
$\tilde{\Csc}_\Mc \equiv n\, n!\left( N_f N_c N_D\right)^{\f{n}{2}}
\left(\f{\Csc_\Mc}{2N_f}\right)^{\f{1}{2}} $. I now make the
assumptions that $\Vert \propa_M \Vert\le m_0^{-1} < \infty$ and
$\Vert \deltaD\Vert\le \ded<\infty$, with $m_0$ and $\ded$ independent
of the gauge configuration.  With these assumptions one concludes
\begin{equation}
  \label{eq:bound_summary2}
  \begin{aligned}
    \left|\left\la \Cc^a \Oc_\Mc\right\ra\right| &= \left|
      \f{\de}{\de\theta_a}\left\la \Oc_\Mc^{\bm{\theta}} \right\ra
      \Big|_{\bm{\theta}=0} \right| \le \delta m \f{\tilde{\Csc}_\Mc
      \ded}{m_0^{n+1}}\,.
  \end{aligned}
\end{equation}
Using also the assumption that $m_0$ and $\ded$ are independent of the
lattice size $\Vc$, the bound Eq.~\eqref{eq:bound_summary2} is
volume-independent and therefore holds also in the thermodynamic
limit; using the continuity in mass of $\Csc_\Mc$ (or its boundedness
near the symmetric point) and the assumed finiteness of $m_0$ and
$\ded$ in the symmetric limit one concludes that
\begin{equation}
  \label{eq:bound_summary3}
  \begin{aligned}
    & \lim_{\delta m\to 0}\lim_{\Vc\to\infty}
    \f{\de}{\de\theta_a}\left\la \Oc_\Mc^{\bm{\theta}} \right\ra
    \Big|_{\bm{\theta}=0} = 0\,,
\end{aligned}
\end{equation}
which is what had to be proved [see Eq.~\eqref{eq:task_inf}].

Notice that under the same assumptions used above one can show that
the expectation values $\left\la \Oc_\Mc \right\ra$ are indeed finite
in the thermodynamic limit, independently of the choice of masses. In
fact, a bound similar to Eq.~\eqref{eq:cs1} is obtained for
$\la \Oc_\Mc \ra$ starting from Eq.~\eqref{eq:gen_obs_F},
\begin{equation}
  \label{eq:cs0}
  \begin{aligned}
    \left|\left\la \Oc_\Mc \right\ra\right| & \le \sum_{\mathrm{P}\in
      \mathrm{S}_{n}} \left\la\left( \Kc_\Mc(\colx,\coly)
        \prod_{k=1}^{n} \Sc({y}_k,{x}_{\mathrm{P}(k)})
      \right)^{\f{1}{2}}\right\ra_{\mathrm{G}} \,.
  \end{aligned}
\end{equation}
Under the boundedness assumptions on $\Vert \propa_M\Vert$ and the
continuity assumption on $\Mc$ (or the boundedness assumption on $\Kc_\Mc$) 
\begin{equation}
  \label{eq:cs0_1}
  \begin{aligned}
    \left|\left\la \Oc_\Mc \right\ra\right| & \le n!\,\left(N_f N_c
      N_D\right)^{\f{n}{2}} \left\la
      \Kc_\Mc(\colx,\coly)^{\f{1}{2}}\Vert
      \propa_M\Vert^{n}\right\ra_{\mathrm{G}} \\ & \le
    \f{n!\,\left(N_f N_c
        N_D\right)^{\f{n}{2}}\Csc_\Mc^{\f{1}{2}}}{m_0^n} \,,
  \end{aligned}
\end{equation}
which is a finite bound, independent of $\Vc$. The extension of this
result and of Eq.~\eqref{eq:bound_summary3} to the case of absolutely
integrable $\Mc_{\colA\colB}$ is discussed in Appendix
\ref{sec:abs_int}.

\section{Application to specific discretizations}  
\label{sec:examples}

In this Section I discuss explicitly several lattice discretizations
of the single-flavor Dirac operator. A superscript is used to
distinguish them and the corresponding propagators, i.e.,
$D^{\mathrm{X}(m)}=\dmlgen+m\deltaD^{\mathrm{X}}$,
$\propa_M^{\mathrm{X}}=\diag(S^{\mathrm{X}(m_1)},\ldots,
S^{\mathrm{X}(m_{N_f})})$,
$S^{\mathrm{X}(m)}=(D^{\mathrm{X}(m)})^{-1}$.  I discuss in particular
staggered fermions (S), Ginsparg-Wilson fermions (GW), Wilson fermions
(W), and minimally doubled fermions (KW, BC), on hypercubic lattices.
Lattice sites are labeled by coordinates $x_\mu=0,\ldots, L_\mu-1$,
where $L_\mu$ is the linear size in direction $\mu$, with
$\mu=1,\ldots, d$. The lattice oriented edges connect $x$ and
$x+\hat{\mu}$, with $\hat{\mu}$ the unit vector in direction $\mu$;
the associated link variables are denoted with $U(x,x+\hat{\mu})$, and
$U(x,x-\hat{\mu})\equiv U(x-\hat{\mu},x)^\dag$ denotes the link
variable associated with the oppositely oriented edge. Finally,
$\delta(x,y) = \prod_{\mu=1}^{d}\delta_{x_\mu\,y_\mu}$. Standard
boundary conditions (periodic for link variables and
periodic/antiperiodic in space/time for Grassmann variables) are
understood, although they do not play any particular role.

\subsection{Staggered fermions}
\label{sec:sf}

The case of staggered fermions~\cite{Kogut:1974ag,Susskind:1976jm,
  Banks:1976ia} is the most straightforward. The corresponding
discretization of the lattice Dirac operator carries no Dirac index,
and reads
\begin{equation}
  \label{eq:stag}
  \begin{aligned}
    \dmls(x,y)&= \tf{1}{2}\textstyle{\sum_{\mu=1}^d} U(x,y)
    \left(\eta_\mu(x)\delta(x+\hat{\mu},y) \right. \\ &
    \phantom{=\f{1}{2}\textstyle{\sum_{\mu=1}^d} U(x,y)}\left.  -
      \eta_\mu(y) \delta(x-\hat{\mu},y)\right)
    \,,\\
    \deltaD^{\mathrm{S}}(x,y)&= \mathbf{1}_C\delta(x,y) \,,
  \end{aligned}
\end{equation}
where $\eta_\mu(x)=(-1)^{\sum_{\alpha<\mu}x_\alpha}$. Notice that
$L_\mu$ must be an even number for every $\mu$. The staggered operator
is anti-Hermitean, and obviously commutes with
$\deltaD^{\mathrm{S}}$. Let $i\lambda_n$, $\lambda_n\in\mathbb{R}$, be
its purely imaginary eigenvalues. Since $\dmls$ has the chiral
property $\{\varepsilon,\dmls\}=0$, where
$\varepsilon_{ab}(x,y)=(-1)^{\sum_\alpha
  x_\alpha}\delta_{ab}\delta(x,y)$, these come in complex conjugate
pairs $\pm i\lambda_n$ or vanish, and so
\begin{equation}
  \label{eq:detstag}
  \det
  D^{\mathrm{S}(m)}=m^{\Nc_0^{\mathrm{S}}}\prod_{n,\,\lambda_n>0}(\lambda_n^2
  +  m^2)\,, 
\end{equation}
where $\Nc_0^{\mathrm{S}}$ is the number of exact zero modes, which
must be an even number. The integration measure $d\mu_{\mathrm{G}}$ is
therefore positive for any choice of $m_f$. For the propagator one has
 \begin{equation}
  \label{eq:stag2_0}
  \begin{aligned}
    S^{\mathrm{S}(m)\,\dag}S^{\mathrm{S}(m)}&= \left(m^2-\dmls{}^{\,2}
    \right)^{-1}\,,
    \\
    \Vert S^{\mathrm{S}(m)}\Vert^2 &= \f{1}{m^2 + \min_n\lambda_n^2}
    \le \f{1}{m^2}\,,
  \end{aligned}
\end{equation}
and so
\begin{equation}
  \label{eq:stag2}
  \begin{aligned}
    \Vert \propa_M^{\mathrm{S}}\Vert^2 & = \max_f \Vert
    S^{\mathrm{S}(m_f)}\Vert^2 = \f{1}{\min_fm_f^2 +
      \min_n\lambda_n^2} \\ & \le \f{1}{\min_f m_f^2} \equiv
    \f{1}{m_0^2}\,.
  \end{aligned}
\end{equation}
Obviously $\Vert\deltaD^{\mathrm{S}}\Vert =1$.  All the assumptions
used in Section \ref{sec:bounds} hold, and so vector flavor symmetry
cannot be spontaneously broken, as long as the common fermion mass in
the symmetric limit is nonzero. This was already
known~\cite{Aloisio:2000rb}. The result holds at any lattice spacing,
and remains true for any choice of boundary conditions or inclusion of
external fields (e.g., an imaginary chemical potential or a magnetic
field) that preserves the anti-Hermiticity and the chiral property of
$\dmls$.

The result still holds also for improved staggered operators as long
as they retain these properties. In particular, this is the case if in
Eq.~\eqref{eq:stag} one replaces the ``thin links'' $U$ with ``fat
links'' obtained by some smearing procedure; and if one improves the
lattice approximation of the covariant derivative by including terms
that only couple even and odd lattice sites (i.e., sites with
$\sum_\alpha x_\alpha$ even or odd), e.g., the Naik
term~\cite{Naik:1986bn}. This covers all the commonly used improved
operators (e.g., ASQTAD~\cite{Lepage:1998vj}, stout
smeared~\cite{Morningstar:2003gk}, HISQ~\cite{Follana:2006rc}).

Extending the result to rooted staggered fermions is not entirely
straightforward. The expectation values of localized observables
$\Oc_\Mc$ in the rooted theory with $N_f$ flavors of staggered quarks
are obtained by replacing the fermionic determinant for each flavor
with its positive fourth root (so keeping the integration measure
positive), and by including suitable counting factors for the various
permutations appearing in the fermionic expectation value
$\la \Oc_\Mc\ra_{\mathrm{F}}$, Eq.~\eqref{eq:gen_obs_F}.  Defining
vector flavor transformations on these observables directly as
transformations of $\Mc$, i.e.,
$\Oc_\Mc\to \Oc_\Mc^{\bm{\theta}}=\Oc_{\Mc^{\bm{\theta}}}$ [see
Eq.~\eqref{eq:wi1bis_app2} and comment after
Eq.~\eqref{eq:wi11quater}], one can still show that the variation of
$\la\Oc_\Mc\ra$ for infinitesimal $\bm{\theta}$ is bounded by the mass
spread $\delta m$ times a constant, hence it vanishes in the symmetric
limit. In fact, such a variation remains of the form
Eq.~\eqref{eq:wi11quater} up to inclusion of the counting factors; the
rest of the argument is unchanged.  However, one should still check
that the flavor transformations defined above reduce in the continuum
limit to the correct transformations of the physical subset of
fermionic degrees of freedom. To this end, one may use the blocking
transformations and the reweighted actions of
Refs.~\cite{Shamir:2004zc,Shamir:2006nj}, introduced to argue the
validity of the rooting procedure (I thank an anonymous referee for
pointing these references out). Such an analysis is, however, beyond
the scope of this paper.  It should be noted that if one accepts the
validity of the rooting procedure, then the uniform bound of
Ref.~\cite{Aloisio:2000rb} on the staggered propagator in a gauge
field background suffices to prove the absence of massless particles
in the spectrum of the continuum theory for any nonzero common fermion
mass, implying the impossibility of spontaneous flavor symmetry
breaking.

\subsection{Ginsparg-Wilson fermions}
\label{sec:gwf}

Massless Ginsparg-Wilson fermions are characterized by the
relation~\cite{Ginsparg:1981bj}
\begin{equation}
  \label{eq:gw}
  \{  \dmlgw,\gamma_5\} = 2\dmlgw R\gamma_5 \dmlgw\,,
\end{equation}
with $R$ a local operator, satisfied by the corresponding lattice
discretization $\dmlgw$ of the Dirac operator. Most of the known
examples~\cite{Hasenfratz:1993sp,DeGrand:1995ji,Hasenfratz:1998ri,
  Kaplan:1992bt,Shamir:1993zy,Narayanan:1993sk,Narayanan:1993ss,
  Neuberger:1997fp,Neuberger:1998wv} satisfy this relation with
$2R=\mathbf{1}$, and moreover are $\gamma_5$-Hermitean,
\begin{equation}
  \label{eq:gw2}
  \gamma_5 \dmlgw\gamma_5 = \dmlgw{}^{\,\dag}\,.
\end{equation}
If these extra assumptions hold it is easy to show that 
\begin{equation}
  \label{eq:gw3}
  \begin{aligned}
    (\dmlgw-\mathbf{1})( \dmlgw-\mathbf{1})^\dag &=\mathbf{1}\,,
  \end{aligned}
\end{equation}
i.e., $\dmlgw= \mathbf{1}+ \Uc$ with $\Uc$ unitary. For massive
Ginsparg-Wilson fermions one uses
\begin{equation}
  \label{eq:gw4}
  \deltaD^{\mathrm{GW}} = \mathbf{1} - \tf{1}{2}\dmlgw\,,
\end{equation}
and so
\begin{equation}
  \label{eq:gw5}
  D^{\mathrm{GW}(m)}= \left(1+\tf{m}{2}\right)\mathbf{1} +
  \left(1-\tf{m}{2}\right)\Uc\,. 
\end{equation}
This is a normal operator with spectrum lying on a circle of radius
$|1-\f{m}{2}|$ centered at $1+\f{m}{2}$, so its eigenvalues are
bounded in magnitude from below by the square root of
\begin{equation}
  \label{eq:gw6}
  \begin{aligned}
    & \min_\varphi\left| \left(1+\tf{m}{2}\right) +
      \left(1-\tf{m}{2}\right)e^{i\varphi}\right|^2 = \min(m^2,4)\,.
  \end{aligned}
\end{equation}
It follows that the propagator obeys
$\Vert \propa^{\mathrm{GW}}_M \Vert \le \max \left(\f{1}{\min_f
    |m_f|},\f{1}{2}\right)$, which is a finite bound if $m_f\neq 0$
$\forall f$.  For $\deltaD^{\mathrm{GW}}$ one has
\begin{equation}
  \label{eq:gw7}
  \begin{aligned}
    &\Vert\deltaD^{\mathrm{GW}}\Vert^2 \le
    \tf{1}{2}\max_\varphi\left|1 - e^{i\varphi}\right|^2 =2 \equiv
    \ded^2 <\infty
    \,.
  \end{aligned}
\end{equation}
As a consequence of Eq.~\eqref{eq:gw2},
$\gamma_5 \Uc \gamma_5 = \Uc^\dag$, and so if $\psi_n$ is a common
eigenvector of $\dmlgw$ and $\dmlgw{}^{\,\dag}$ with eigenvalues
$\mu_n=1+e^{i\varphi_n}$ and $\mu_n^*= 1+e^{-i\varphi_n}$,
respectively, then
$\dmlgw\gamma_5\psi_n =\gamma_5 \dmlgw{}^{\,\dag} \psi_n =
\mu_n^*\gamma_5 \psi_n$. It follows that complex eigenvalues come in
complex-conjugate pairs; for the real eigenvalues $\mu_n=\mu_n^*=0,2$,
one can instead choose chiral eigenvectors $\psi_\pm$, satisfying
$\gamma_5\psi_\pm=\pm\psi_\pm$. For the determinant of
$D^{\mathrm{GW}(m)}$ one finds
\begin{equation}
  \label{eq:detGW}
  \begin{aligned}
    \det D^{\mathrm{GW}(m)} & = m^{\Nc_0^{\mathrm{GW}}}
    2^{\Nc_2^{\mathrm{GW}}} \\ &\phantom{=}\times
    \prod_{n,\,\sin\varphi_n>0}
    \left[\left(2\cos\tf{\varphi_n}{2}\right)^2 +
      \left(m\sin\tf{\varphi_n}{2}\right)^2 \right]\,,
  \end{aligned}
\end{equation}
with $\Nc_{0,2}^{\mathrm{GW}}$ the degeneracies of the two real
eigenvalues.  It follows that the integration measure
$d\mu_{\mathrm{G}}$ is positive if $m_f\ge 0$ $\forall f$, and more
generally for an even number of negative masses. Vector flavor
symmetry cannot be spontaneously broken in the symmetric limit as long
as the common fermion mass is positive, or just nonzero if $N_f$ is
even, at any value of the lattice spacing. The use of different
boundary conditions or the inclusion of external fields in $\dmlgw$
does not change this result, as long as the operator remains of the
form $\dmlgw=\mathbf{1}+\Uc$ and the $\gamma_5$-Hermiticity property
Eq.~\eqref{eq:gw2} holds.

\subsection{Wilson fermions}
\label{sec:wf}

For Wilson fermions~\cite{Wilson:1975id} the massless operator
$\dmlw=\dmln+R^{\mathrm{W}}$ is obtained adding the naive
discretization $\dmln$ of the massless Dirac operator and the Wilson
term $R^{\mathrm{W}}$, while $\deltaD^{\mathrm{W}}$ is the identity
in color, Dirac, and coordinate space,
\begin{equation}
  \label{eq:wilson}
  \begin{aligned}
    \dmln(x,y) &= \tf{1}{2}\textstyle{\sum_{\mu=1}^d} U(x,y)\gamma_\mu
    \left(\delta(x+\hat{\mu},y)\right. \\
    &\phantom{=\tf{1}{2}\textstyle{\sum_{\mu=1}^d}
      U(x,y)\gamma_\mu()}\left.  - \delta(x-\hat{\mu},y)
    \right)\,,\\
    R^{\mathrm{W}}(x,y) &= -\tf{r}{2}\mathbf{1}_D
    \textstyle{\sum_{\mu=1}^d}
    \left(  U(x,y) \left( \delta(x+\hat{\mu},y) \right. \right.\\
    &\phantom{= -\tf{r}{2}\mathbf{1}_D
      \textstyle{\sum_{\mu=1}^d}U(x,y)()}\left.\left.+
        \delta(x-\hat{\mu},y)\right)
    \right. \\
    &\phantom{= -\tf{r}{2}\mathbf{1}_D
      \textstyle{\sum_{\mu=1}^d}()}\left.-
      2\mathbf{1}_{C}\delta(x,y)\right)\,,\\
    \deltaD^{\mathrm{W}}(x,y)
    &=\mathbf{1}_{C}\mathbf{1}_{D}\delta(x,y)\,,
  \end{aligned}
\end{equation}
with $r$ a nonzero real parameter. This operator is not anti-Hermitean
and not even normal, satisfying only the $\gamma_5$-Hermiticity
condition
$\gamma_5 D^{\mathrm{W}(m)}\gamma_5 = D^{\mathrm{W}(m)\,\dag}$. The
spectrum of $D^{\mathrm{W}(m)}$ is generally complex, and while
$\gamma_5$-Hermiticity guarantees that $\det D^{\mathrm{W}(m)}$ is
real, one is not guaranteed to find a positive integration measure
$d\mu_{\mathrm{G}}$, unless an even number of fermions with the same
mass is present. Moreover, while $\deltaD^{\mathrm{W}}$ is obviously
bounded, no general lower bound applies to the spectrum of
$D^{\mathrm{W}(m)\,\dag}D^{\mathrm{W}(m)}$, even in the massive case,
and so no uniform upper bound on the norm of the propagator is
available. The result of the previous Section therefore does not apply
to Wilson fermions. This is not surprising since it is known that
vector flavor symmetry is spontaneously broken in the Aoki
phase~\cite{Aoki:1983qi,Aoki:1986xr,Setoodeh:1988ds,
  Aoki:1989rw,Aoki:1990ap,Aoki:1992nb,Aoki:1995yf,Aoki:1995ft,
  Aoki:1995ba,Aoki:1996pw,Aoki:1996af,Aoki:1997fm,Bitar:1996kc,
  Bitar:1997as,Bitar:1997ic,Edwards:1997sp,Edwards:1998sh,
  Sharpe:1998xm,Azcoiti:2008dn,Sharpe:2008ke,Azcoiti:2012ns}. One can,
however, refine the discussion and see more precisely how things fail
for Wilson fermions.

Presumably, for an even number of flavors and sufficiently small
$\delta m$ the sign problem of the integration measure affects only a
set of gauge configurations of zero measure. If so, in this case the
integration measure would effectively be positive, and so one could
follow the derivation of the previous Section up to
Eq.~\eqref{eq:bound_summary}, obtaining
\begin{equation}
  \label{eq:bound_summary_Wilson}
  \begin{aligned}
    \left| \f{\de}{\de\theta_a}\left\la \Oc_\Mc^{\bm{\theta}}
      \right\ra \Big|_{\theta=0} \right| &\le \delta m\,
    \tilde{\Csc}_\Mc \left\la \Vert \propa_M^{\mathrm{W}} \Vert^{n+1}
    \right \ra_{\mathrm{G}}\\ & = \delta m\, \tilde{\Csc}_\Mc
    \int_0^\infty ds\,p(s) s^{n+1} \,,
  \end{aligned}
\end{equation}
where
\begin{equation}
  \label{eq:wilson2}
  p(s)\equiv  \left\la \delta\left(
      s-  \Vert  \propa_M^{\mathrm{W}}
      \Vert    \right)
  \right \ra_{\mathrm{G}}\,.
\end{equation}
One could then still exclude the spontaneous breaking of vector flavor
symmetry if $p(s)$ vanished faster than any polynomial as
$s\to\infty$, for example if $p(s)=0$ for $s> s_0$ for some $s_0$, or
if it vanished exponentially. If $p(s)$ vanished only as a power law
$p(s)\sim s^{-n_0}$, or not at all, the argument above would not
provide a viable bound for $n >n_0-2$, and spontaneous breaking could
not be excluded. In fact, in the Aoki phase one finds a finite
spectral density of near-zero modes of the Hermitean operator
$H=\gamma_5D^{\mathrm{W}(m)}$~\cite{Setoodeh:1988ds,Edwards:1997sp,
  Edwards:1998sh}. Since
\begin{equation}
  \label{eq:gamma5W}
  \begin{aligned}
    \Vert \propa^{\mathrm{W}(m)}\Vert^2 &= \sup_{\psi\neq 0}\f{(\psi,
      \left(D^{\mathrm{W}(m)}D^{\mathrm{W}(m)\,\dag}\right)^{-1}
      \psi)}{(\psi,\psi)} \\ &= \sup_{\psi\neq 0}\f{(\gamma_5\psi,
      \left(H^2\right)^{-1}\gamma_5
      \psi)}{(\gamma_5\psi,\gamma_5\psi)}= \f{1}{\min_n h_n^2} \,,
  \end{aligned}
\end{equation}
where $h_n\in\mathbb{R}$ are the eigenvalues of $H$, one finds that as
the lattice volume grows and the lowest mode of $H$ on typical
configurations tends to zero $p(s)$ becomes more and more peaked at a
larger and larger value of $s$, eventually tending to infinity in the
thermodynamic limit, and the right-hand side of
Eq.~\eqref{eq:bound_summary_Wilson} blows up, making the bound
useless. Outside the Aoki phase the spectrum of $H$ is gapped around
the origin, the right-hand side of Eq.~\eqref{eq:bound_summary_Wilson}
is finite, and the bound prevents spontaneous flavor symmetry
breaking.

\subsection{Minimally doubled fermions}
\label{sec:mindoub}

For the minimally doubled fermions of Karsten and Wilczek
(KW)~\cite{Karsten:1981gd,Wilczek:1987kw}, and of Creutz and Bori{\c
  c}i (BC)~\cite{Creutz:2007af,Borici:2007kz}, the massless Dirac
operator is of the form $ \dmlmd = \dmln + R^{\mathrm{X}}$,
$ \mathrm{X}=\mathrm{KW},\mathrm{BC}$, where the naive operator
$\dmln$ is defined in Eq.~\eqref{eq:wilson}, and the inclusion of the
terms
\begin{equation}
  \label{eq:kwbc2}
    \begin{aligned}
      R^{\mathrm{KW}}(x,y) &= - \tf{ir}{2}
      \gamma_{d}\textstyle{\sum_{\mu=1}^{d-1}}
      \left(U(x,y)\left(\delta(x+\hat{\mu},y) \right.\right.\\
      &\phantom{=- \tf{ir}{2}
        \gamma_{d}\textstyle{\sum_{\mu=1}^{d-1}}()U(x,y)}\left.\left.
          + \delta(x-\hat{\mu},y)\right) \right.\\ &\phantom{=-
        \tf{ir}{2} \gamma_{d}\textstyle{\sum_{\mu=1}^{d-1}}()}\left.
        - 2\mathbf{1}_{C}\delta(x,y) \right) \,,
      \\
      R^{\mathrm{BC}}(x,y) &= - \tf{ir}{2} \textstyle{\sum_{\mu=1}^d}
      \gamma_\mu' \left(U(x,y)\left(\delta(x+\hat{\mu},y)
        \right.\right.\\ &\phantom{=- \tf{ir}{2}
        \textstyle{\sum_{\mu=1}^{d}}()U(x,y)\gamma_\mu'}\left.\left.
          + \delta(x-\hat{\mu},y)\right) \right.\\ &\phantom{=-
        \tf{ir}{2} \textstyle{\sum_{\mu=1}^{d}}()\gamma_\mu'}\left.  -
        2\mathbf{1}_{C}\delta(x,y)\right) \,,
    \end{aligned}
\end{equation}
where $\gamma_\mu' \equiv \Gamma\gamma_\mu\Gamma$ and
$\Gamma\equiv \f{1}{\sqrt{d}}\sum_{\nu=1}^{d}\gamma_\nu$, reduces the
number of doublers to two when $r=1$.  The massive operator is
obtained in both cases using the trivial mass term
$\deltaD^{\mathrm{KW}}=\deltaD^{\mathrm{BC}}= \deltaD^{\mathrm{W}}$,
see again Eq.~\eqref{eq:wilson}. For both types of fermions the
massless operator is anti-Hermitean and chiral,
$\{\dmlmd,\gamma_5\}=0$, and obviously commutes with the mass term.
One can then diagonalize $\dmlmd$ obtaining purely imaginary
eigenvalues $i\lambda_n^{\mathrm{X}}$ and a symmetric spectrum, and so
the single-flavor propagators $S^{\mathrm{X}(m)}$ obey
\begin{equation}
  \label{eq:kwbc3}
  \Vert S^{\mathrm{X}(m)} \Vert^2 = \f{1}{m^2 + \min_n
    \lambda_n^{\mathrm{X}\,2}}\le 
  \f{1}{m^2}\,,\qquad
  \mathrm{X}=\mathrm{KW},\mathrm{BC}\,.
\end{equation}
The fermionic determinant reads 
\begin{equation}
  \label{eq:detmindoub}
  \det
  D^{\mathrm{X}(m)}=m^{\Nc_0^{\mathrm{X}}}\prod_{n,\,\lambda_n>0}(\lambda_n^{\mathrm{X}\,2}
  +   m^2)\,,\qquad
  \mathrm{X}=\mathrm{KW},\mathrm{BC}\,,
\end{equation}
with $\Nc_0^{\mathrm{X}}$ the number of exact zero modes, so it is
positive for nonnegative fermion masses, and for an even number of
negative masses. The same argument therefore applies as with staggered
fermions, and vector flavor symmetry cannot break spontaneously as
long as the common fermion mass is positive (or just nonzero if $N_f$
is even) in the symmetric limit, independently of the lattice spacing.

\section{Conclusions}
\label{sec:concl}

In this paper I have shown that under quite general assumptions on the
discretization of the Dirac operator $\Dirac_M$, one can rigorously
exclude the possibility of spontaneous breaking of vector flavor
symmetry on the lattice, in gauge theories with a positive
functional-integral measure. These assumptions are
\begin{enumerate}
  \setcounter{enumi}{-1}
\item $\Dirac_M$ is linear in the fermion masses,
  $\Dirac_M=\dml + M\deltaD$, with $\dml$ and $\deltaD$ trivial in
  flavor space, and $M$ a Hermitean mass matrix;
\item the norm of the propagator $\Dirac_M^{-1}$ can be bounded by a
  configuration- and volume-in\-de\-pen\-dent quantity, that remains
  finite in the symmetric limit of fermions of equal masses,
  $M\to m\mathbf{1}_F$;
\item the norm of the derivative of $\Dirac_M$ with respect to the
  fermion masses, $\deltaD$, can be bounded by a configuration- and
  volume-independent quantity, that remains finite as
  $M\to m\mathbf{1}_F$.
\end{enumerate}
The impossibility of spontaneous flavor symmetry breaking on the
lattice is proved by showing that any localized order parameter must
vanish in the symmetric limit, taken after the thermodynamic limit. If
the assumptions above hold for any (or at least for sufficiently
small) lattice spacing, this result remains true also in the continuum
limit, if this exists. My argument applies in particular to staggered
fermions~\cite{Kogut:1974ag,Susskind:1976jm,Banks:1976ia}; to the
minimally doubled fermions of Karsten and
Wilczek~\cite{Karsten:1981gd,Wilczek:1987kw} and of Creutz and Bori{\c
  c}i~\cite{Creutz:2007af,Borici:2007kz}; and to Ginsparg-Wilson
fermions~\cite{Ginsparg:1981bj,Hasenfratz:1993sp,DeGrand:1995ji,
  Hasenfratz:1998ri,Kaplan:1992bt,Shamir:1993zy,Narayanan:1993sk,
  Narayanan:1993ss,Neuberger:1997fp,Neuberger:1998wv} that are
$\gamma_5$-Hermitean and satisfy the Ginsparg-Wilson relation with
$2R=\mathbf{1}$ [see Eq.~\eqref{eq:gw}]. For these discretizations one
can exclude spontaneous breaking of vector flavor symmetry on the
lattice for any spacing, and so in the continuum limit as well, for
any positive common fermion mass $m$ (and for any nonzero $m$ for an
even number of flavors). Quite unsurprisingly, the argument fails in
the case of Wilson fermions~\cite{Wilson:1975id}, where such a
spontaneous breaking is known to happen in the Aoki
phase~\cite{Aoki:1983qi,Aoki:1986xr,Setoodeh:1988ds,
  Aoki:1989rw,Aoki:1990ap,Aoki:1992nb,Aoki:1995yf,Aoki:1995ft,
  Aoki:1995ba,Aoki:1996pw,Aoki:1996af,Aoki:1997fm,Bitar:1996kc,
  Bitar:1997as,Bitar:1997ic,Edwards:1997sp,Edwards:1998sh,
  Sharpe:1998xm,Azcoiti:2008dn,Sharpe:2008ke,Azcoiti:2012ns}. While
for staggered fermions spontaneous breaking of vector flavor symmetry
(as well as of baryon number symmetry) was already completely excluded
by the results of Ref.~\cite{Aloisio:2000rb}, for Ginsparg-Wilson
fermions only partial results were previously
available~\cite{Azcoiti:2010ns}.

My result is clearly not as powerful as that obtained by Vafa and
Witten working with the continuum functional integral in
Ref.~\cite{Vafa:1983tf}, and by Aloisio \textit{et al.}\ in
Ref.~\cite{Aloisio:2000rb} working with staggered fermions on the
lattice. In particular, although it excludes the possibility of
Goldstone bosons appearing in the spectrum due to spontaneous flavor
symmetry breaking, it cannot exclude completely the presence of
massless bosons, as Refs.~\cite{Vafa:1983tf,Aloisio:2000rb} do.  On
the other hand, the use of a properly regularized functional integral
rather than the continuum one used in Ref.~\cite{Vafa:1983tf} makes
the present argument mathematically fully rigorous. Since the
ultralocality and anti-Hermiticity of the staggered operator are not
used, as they are in Ref.~\cite{Aloisio:2000rb}, my argument works
also for more general discretizations, in particular allowing one to
treat the case of an arbitrary number of physical fermion flavors in
the continuum limit without resorting to the ``rooting trick.''

The bound on the variation of expectation values under a vector flavor
transformation [see Eq.~\eqref{eq:bound_summary2}] proved here to
derive the main result is probably far less than optimal, as it does
not take into account the cancellations present in fermionic
observables due to the oscillating sign of the contributions of the
various field contractions. The bound on the propagator [see
Eq.~\eqref{eq:bound_prop_good3}] is also likely to be suboptimal, and
one suspects that a lattice analogue of the Vafa-Witten bound could be
obtained also for more general discretizations than staggered
fermions, for which it was proved in Ref.~\cite{Aloisio:2000rb}. A
direct extension of the proof of Ref.~\cite{Aloisio:2000rb} to
minimally doubled fermions seems feasible, while a different approach
is probably needed for Ginsparg-Wilson fermions. It is worth noting,
however, that a global, coordinate-independent bound like
Eq.~\eqref{eq:bound_prop_good3} suffices to prove the impossibility of
vector flavor symmetry breaking, without the need to bound the
long-distance behavior of the propagator as in
Refs.~\cite{Vafa:1983tf,Aloisio:2000rb}.

The present argument does not rule out the appearance of phases with
spontaneously broken vector flavor symmetry on the lattice if terms of
order higher than quadratic are included in the fermionic action, even
if the quadratic terms satisfy assumptions (0.)--(2.). Non-symmetric
vacua may in fact exist, degenerate with the symmetric one in the
symmetric limit, but with ground state energy increased by the
standard symmetry-breaking term used here.  These vacua could not be
reached with the procedure used here, and would require the addition
of different symmetry-breaking terms to the symmetric action in order
to select them. This possibility is of limited interest in the
physical case of QCD, since in this theory vector flavor symmetry is
broken explicitly precisely by the differences in the quark masses,
and the symmetric limit of interest where one should investigate the
possibility of its spontaneous breaking is the one considered in this
paper.  More generally, while such spontaneously broken phases on the
lattice could be problematic for numerical simulations, they should be
unphysical and not survive the continuum limit.

The restriction to a quadratic lattice action is in fact not really a
limitation as far as the usual continuum limit is concerned. For
continuum gauge theories in dimension $d>2$ (in $d\le 2$ the
spontaneous breaking of a continuous symmetry is
forbidden~\cite{Mermin:1966fe,Hohenberg:1967zz,Coleman:1973ci}) there
are no perturbatively renormalizable fermionic operators with the
right global and local symmetries other than the quadratic ones,
approximated on the lattice by the action used here. The inclusion of
higher order terms in the lattice action only adds perturbatively
non-renormalizable interactions that do not affect the long-distance
physics in the usual continuum limit. Hypothetical spontaneously
broken phases on the lattice should then shrink as the continuum limit
is approached, with vector flavor symmetry being realized in the
continuum theory. Phases with spontaneously broken vector flavor
symmetry may still be found in the continuum if unconventional
continuum limits exist, but this would concern a different type of
continuum theories. Universality of the continuum limit also implies
that the spontaneously broken phases potentially found on the lattice
for quadratic actions not satisfying assumptions (0.)--(2.)\ should
shrink in the usual continuum limit, as is the case for the Aoki phase
of Wilson fermions.

In conclusion, the existence of lattice discretizations of the Dirac
operator, free of doublers, for which spontaneous vector flavor
symmetry breaking for finite positive fermion mass is impossible at
any lattice spacing (i.e., the Ginsparg-Wilson fermions discussed
above) implies the same impossibility in the continuum limit, if this
exists, for an arbitrary number of fermion species. This settles the
issue of spontaneous vector flavor symmetry breaking in a rigorous
manner (for a physicist's standard of rigor).

\begin{acknowledgments}
  I thank C.~Bonati and M.~D'Elia for discussions, and V.~Azcoiti for
  discussions and a careful reading of the manuscript. This work was
  partially supported by the NKFIH grant KKP-126769.
\end{acknowledgments}

\appendix

\section{Technical details} 
\label{sec:app}

\subsection{Integrated Ward-Takahashi identity and finite flavor
  transformations}
\label{sec:finflavtr}

Since the Berezin integration measure is invariant under vector flavor
transformations, Eq.~\eqref{eq:wi0}, one has after changing variables
[see Eq.~\eqref{eq:w1_not} for the notation]
\begin{widetext}
\begin{equation}
  \label{eq:wi1}
  \begin{aligned}
    \int [\fmeas] \,e^{-\act_{\mathrm{F}}[\psi,\bar{\psi},U]}
    \Oc^{\bm{\theta}}[\psi,\bar{\psi},U] &= \int [\fmeas]
    \,e^{-\bar{\psi}V(\bm{\theta})\Dirac_M[U]
      V(\bm{\theta})^\dag\psi}\Oc[\psi,\bar{\psi},U]\,.
  \end{aligned}
\end{equation}
For infinitesimal $\theta_a$, expanding both sides of the equation to
leading order in $\theta_a$, one finds
\begin{equation}
  \label{eq:wi2}
  \begin{aligned}
    & \int [\fmeas]
    \,e^{-\act_{\mathrm{F}}[\psi,\bar{\psi},U]}\left(\Oc[\psi,\bar{\psi},U]
      + \sum_a\theta_a\f{\de}{\de\theta_a}
      \Oc^{\bm{\theta}}[\psi,\bar{\psi},U]\Big|_{\bm{\theta}=0}+O(\theta^2)\right)
    \\ &= \int [\fmeas] \,e^{-\act_{\mathrm{F}}[\psi,\bar{\psi},U]}
    \left(\Oc[\psi,\bar{\psi},U] - i\left(\bar{\psi}[\bm{\theta}\cdot
        \bm{t}, \Dirac_M[U]]\psi\right) \,\Oc[\psi,\bar{\psi},U] +
      O(\theta^2) \right) \,,
  \end{aligned}
\end{equation}
from which Eq.~\eqref{eq:wi3} follows,
\begin{equation}
  \label{eq:wi3_app}
  \begin{aligned}
    i\f{\de}{\de\theta_a}\left\la \Oc^{\bm{\theta}}\right\ra
    \bigg|_{\bm{\theta}=0} &= \left\la \left(\bar{\psi}[ t^a,
        \Dirac_M]\psi\right) \Oc\right\ra = \left\la \Cc^a
      \Oc\right\ra\,, &&& \Cc^a[\psi,\bar{\psi},U] &\equiv \bar{\psi}[
    t^a, \Dirac_M[U]]\psi\,.
  \end{aligned}
\end{equation}
For finite transformations, since $V(\bm{\theta})$ is analytic in
$\theta_a$ one can write the trivial identity
\begin{equation}
  \label{eq:wi1bis0}
  \begin{aligned}
    & \int [\fmeas] \,e^{-\act_{\mathrm{F}}[\psi,\bar{\psi},U]}
    \left(\Oc^{\bm{\theta}}[\psi,\bar{\psi},U] -\Oc[\psi,\bar{\psi},U]
    \right) = \int_0^1 d\alpha\,\f{d}{d\alpha}\int [\fmeas]
    \,e^{-\act_{\mathrm{F}}[\psi,\bar{\psi},U]}
    \Oc^{\alpha\bm{\theta}}[\psi,\bar{\psi},U]\,.
  \end{aligned}
\end{equation}
Using now Eq.~\eqref{eq:wi1} one finds the following result for the
change of the expectation value of an observable under a finite vector
flavor transformation,
\begin{equation}
  \label{eq:wi1bis}
  \begin{aligned}
    & \int [\fmeas] \,e^{-\act_{\mathrm{F}}[\psi,\bar{\psi},U]}
    \left(\Oc^{\bm{\theta}}[\psi,\bar{\psi},U] -\Oc[\psi,\bar{\psi},U]
    \right) \\
    &= -\int_0^1 d\alpha\, \int [\fmeas]
    \,e^{-\bar{\psi}V(\alpha\bm{\theta})\Dirac_M[U]
      V(\alpha\bm{\theta})^\dag\psi } \left(\bar{\psi}
      V(\alpha\bm{\theta})
      \left[V(\alpha\bm{\theta})^\dag\f{dV(\alpha\bm{\theta})}{d\alpha},
        \Dirac_M[U] \right] V(\alpha\bm{\theta})^\dag\psi\right)\,
    \Oc[\psi,\bar{\psi},U]
    \\
    &= -i\int_0^1 d\alpha\, \int [\fmeas]
    \,e^{-\act_{\mathrm{F}}[\psi,\bar{\psi},U]} \,\left(\bar{\psi}
      \left[ \bm{\theta}\cdot\bm{t} ,\Dirac_M[U] \right] \psi\right)\,
    \Oc^{\alpha\bm{\theta}}[\psi,\bar{\psi},U]\,.
  \end{aligned}
\end{equation}
This implies
\begin{equation}
  \label{eq:wi1bis_app1}
  \begin{aligned}
    \left\la \Oc^{\bm{\theta}} -\Oc \right\ra &= -i\sum_a
    \theta_a\int_0^1 d\alpha\, \left\la \Cc^a\,
      \Oc^{\alpha\bm{\theta}} \right\ra\,.
  \end{aligned}
\end{equation}
For the observables $\Oc_\Mc$ of interest, the effect of a vector
flavor transformation $V(\bm{\theta})$ can be fully accounted for by
replacing $\Mc\to \Mc^{\bm{\theta}}$, i.e.,
$\Oc_\Mc^{\bm{\theta}}=\Oc_{\Mc^{\bm{\theta}}}$, with
\begin{equation}
  \label{eq:wi1bis_app2}
  \Mc^{\bm{\theta}}_{\colA\colB} \equiv
  \Mc_{A_1'\ldots A_n' B_1'\ldots B_n'}
  V(\bm{\theta})^\dag_{A_1 A_1'}\ldots
  V(\bm{\theta})^\dag_{A_n A_n'}
  V(\bm{\theta})_{B_1'B_1}\ldots   V(\bm{\theta})_{B_n'B_n}\,,
\end{equation}
where
$V_{A_i A_i'}=V_{f_i f_i'}\,\delta_{a_i
  a_i'}\delta_{\alpha_i\alpha_i'}$. Notice that
\begin{equation}
  \label{eq:trmmdagtheta}
  \begin{aligned}
    \Kc_{\Mc^{\bm{\theta}}}(\colx,\coly)&=
    \tr_{FCD}\{\Mc^{\bm{\theta}}(\colx,\coly)\Mc^{\bm{\theta}}(\colx,\coly)^\dag\}
    = \tr_{FCD}\{\Mc(\colx,\coly)\Mc(\colx,\coly)^\dag\}=
    \Kc_{\Mc}(\colx,\coly)\,,
  \end{aligned}
\end{equation}
which implies that the same upper bound applies to $\Mc$ and
$\Mc^{\bm{\theta}}$ in Eq.~\eqref{eq:bound_coeff}, i.e.,
$\Csc_{\Mc^{\bm{\theta}}}=\Csc_\Mc$, and the same constant
$\tilde{\Csc}_\Mc=\tilde{\Csc}_{\Mc^{\bm{\theta}}}$ appears in the
bound Eq.~\eqref{eq:bound_summary2}. Using now
Eq.~\eqref{eq:wi1bis_app1} one finds
  \begin{equation}
  \label{eq:wi1bis_app3}
  \begin{aligned}
    \left\la \Oc_\Mc^{\bm{\theta}} -\Oc_\Mc \right\ra &=\left\la
      \Oc_{\Mc^{\bm{\theta}}} -\Oc_\Mc \right\ra = -i\sum_a
    \theta_a\int_0^1 d\alpha\, \left\la \Cc^a \,
      \Oc_{\Mc^{\alpha\bm{\theta}}} \right\ra\,.
  \end{aligned}
\end{equation}
The integrand on the right-hand side obeys the bound
Eq.~\eqref{eq:bound_summary2}, and since
$\tilde{\Csc}_{\Mc^{\alpha\bm{\theta}}}=\tilde{\Csc}_\Mc$ are
independent of $\alpha$ one finds
  \begin{equation}
  \label{eq:wi1bis_app4}
   \begin{aligned}
     \left|\left\la \Oc_\Mc^{\bm{\theta}} -\Oc_\Mc \right\ra \right|
     &\le \sum_a |\theta_a|\int_0^1 d\alpha\, \left|\left\la \Cc^a \,
         \Oc_{\Mc^{\alpha\bm{\theta}}} \right\ra\right| \le \delta m
     \f{\tilde{\Csc}_\Mc\ded }{m_0^{n+1}}\sum_a |\theta_a| \,.
  \end{aligned}
\end{equation}
Since this bound is independent of $\Vc$ and remains finite in the
symmetric limit, one has
  \begin{equation}
  \label{eq:wi1bis_app5}
  \lim_{\delta m \to 0}\lim_{\Vc\to\infty}
  \left\la \Oc_\Mc^{\bm{\theta}} -\Oc_\Mc \right\ra = 0\,. 
\end{equation}
This implies
$\left\la\Oc_\Mc^{\bm{\theta}}\right\ra= \left\la\Oc_\Mc \right\ra $
in the thermodynamic and symmetric limit, for any $\bm{\theta}$, and
so vector flavor symmetry cannot be spontaneously broken if the
assumptions of Section \ref{sec:not} are satisfied.

\subsection{Vanishing of order parameters}
\label{sec:orpar}

Under $\mathrm{SU}(N_f)$ vector flavor transformations, the matrix
$\Mc$ transforms in the product representation with $n$ fundamental
and $n$ antifundamental factors. This is a unitary representation that
can be decomposed in the direct sum of unitary irreducible
representations. The most general $\Mc$ is then the linear combination
of matrices $\Mc^{(R)}_r(\colx,\coly)$,
\begin{equation}
  \label{eq:Mirrep_app}
  \big(\Mc^{(R)}_r\big)_{f_1\Asc_1\ldots f_n \Asc_n g_1\Bsc_1\ldots
    g_n \Bsc_n}(\colx,\coly)= 
  (\Tc^{(R)}_{r})_{f_1\ldots f_n g_1\ldots g_n}
  \Ms_{\Asc_1\ldots\Asc_n\Bsc_1\ldots\Bsc_n}(\colx,\coly) \,,
\end{equation}
where $\Ms$ is a matrix acting only on color and Dirac space that can
depend on the gauge variables in a suitably gauge-covariant way, and
where $\Tc^{(R)}_{r}$, $r=1,\ldots,d_R$, are constant tensors in
flavor space, independent of the gauge variables and of any feature of
the theory other than the number of flavors, that transform
irreducibly under vector flavor transformations [see
Eq.~\eqref{eq:wi1bis_app2} for the notation],
\begin{equation}
  \label{eq:irrtens}
  \Tc^{(R)\,\bm{\theta}}_r 
  = \sum_{r'}\Tc^{(R)}_{r'} D^{(R)}\left(V(\bm{\theta})\right)_{r'r}\,,
\end{equation}
with $D^{(R)}(V)$ the representative of $V\in\mathrm{SU}(N_f)$ in the
irreducible unitary representation $R$ of dimension $d_R$. For these
quantities one has in the thermodynamic and symmetric limit [see
Eq.~\eqref{eq:wi1bis_app5}]
\begin{equation}
  \label{eq:alt_irrep}
    \begin{aligned}
      u^{(R)}_r&\equiv \left\la\Oc_{\Tc^{(R)}_r \Ms} \right\ra =
      \left\la\Oc_{\Tc^{(R)}_r \Ms}^{\bm{\theta}} \right\ra =
      \sum_{r'} \left\la\Oc_{\Tc^{(R)}_{r'} \Ms} \right\ra
      D^{(R)}(V(\bm{\theta}))_{r'r} = \sum_{r'}u^{(R)}_{r'}
      D^{(R)}(V(\bm{\theta}))_{r'r} \,,
 \end{aligned}
\end{equation}
which expresses the invariance of the vector $u^{(R)}$ with components
$u^{(R)}_r$, $r=1\ldots,d_R$, under an arbitrary $\mathrm{SU}(N_f)$
transformation. No nonzero invariant vector exists if $R$ is a
nontrivial representation\footnote{Proof: In matrix notation,
  Eq.~\eqref{eq:alt_irrep} reads $u^{(R)T}= u^{(R)T}D^{(R)}(V)$,
  $\forall V\in\mathrm{SU}(N_f)$, implying also
  $u^{(R)*}=D^{(R)}(V)^\dag u^{(R)*}$. The projector
  $P=u^{(R)*}u^{(R)T}$ is then left invariant by an irreducible
  unitary representation of $\mathrm{SU}(N_f)$,
  $P = D^{(R)}(V)^\dag P D^{(R)}(V)= D^{(R)}(V)^{-1} P D^{(R)}(V)$.
  By Schur's lemma $P$ is then proportional to the $d_R$-dimensional
  identity matrix, but since it has only rank 1 it must vanish if
  $d_R\neq 1$.}, so $u^{(R)}=0$ follows unless $R$ is the
one-dimensional trivial representation, and all order parameters
vanish.

\subsection{Bound on partial traces}
\label{sec:boundpt}

Let $\mathrm{A}$ be a matrix with entries labeled by pairs of an
arbitrary number of indices.  Denote the subset of indices
corresponding to an ``internal'' space $\mathcal{I}$ collectively by
$I$, and denote the remaining ones collectively by $x$, i.e., the
matrix entries are labeled as $\mathrm{A}_{I'I}(x',x)$. The partial
trace over the internal space of the matrix
$\mathrm{A}(x',x)^\dag \mathrm{A}(x',x)$, where Hermitean conjugation
and matrix multiplication are understood to apply only to the internal
indices, satisfies the bound
\begin{equation}
  \label{eq:lemma2}
  \tr_{\mathcal{I}}\mathrm{A}(x',x)^\dag\mathrm{A}(x',x) =
  \tr_{\mathcal{I}}\mathrm{A}(x',x)\mathrm{A}(x',x)^\dag \equiv
  \sum_{I,I'} \mathrm{A}_{I'I}(x',x)\mathrm{A}_{I'I}(x',x)^*  
  \le
  \mathrm{dim}\, \mathcal{I}\,\Vert \mathrm{A} \Vert^2\,,
\end{equation}
where $\Vert\mathrm{A}\Vert$ is the usual operator norm of
$\mathrm{A}$. This is equal to the square root of the largest
eigenvalue of the matrix $\mathrm{A}^\dag \mathrm{A}$, where Hermitean
conjugation and matrix multiplication are understood to apply to all
the indices [see Eqs.~\eqref{eq:matmultexamples} and
\eqref{eq:hermconjgexamples} for the notation].

\paragraph*{Proof} By the spectral theorem, the Hermitean matrix
$\mathrm{A}^\dag \mathrm{A}$ can be written as follows,
\begin{equation}
  \label{eq:lemma_proof}
  \begin{aligned}
    \mathrm{A}^\dag \mathrm{A}&=\sum_n a_n^2 \phi_n\phi_n^\dag \,, &&&
    \left( \mathrm{A}^\dag \mathrm{A}\right)_{I'I}(x',x)&=\sum_n a_n^2
    \phi_{n\,I'}(x')\,\phi_{n\,I}(x)^*\,,
  \end{aligned}
\end{equation}
where $a_n^2$ are its real positive eigenvalues and $\phi_n$ are a
complete set of orthonormal eigenvectors,
$(\mathrm{A}^\dag \mathrm{A})\phi_n=a_n^2\phi_n$, with
\begin{equation}
  \label{eq:lemma_proof2}
  \begin{aligned}
    (\phi_{n'},\phi_{n}) &\equiv
    \sum_{x,I}\phi_{n'\,I}(x)^*\phi_{n\,I}(x) = \delta_{n'n}\,, &&&
    \sum_n\phi_{n\,I'}(x') \phi_{n\,I}(x)^* &=
    \delta_{I'I}\delta(x',x)\,.
  \end{aligned}
\end{equation}
Since $\tr_{\mathcal{I}}\mathrm{A}(x',x)^\dag\mathrm{A}(x',x) \ge 0$
for any $x'$, one finds
\begin{equation}
  \label{eq:lemma_proof3}
  \begin{aligned}
    \tr_{\mathcal{I}}\mathrm{A}(x',x)^\dag\mathrm{A}(x',x) & =
    \tr_{\mathcal{I}}(\mathrm{A}^\dag)(x,x')\mathrm{A}(x',x) \le
    \sum_{x'}\tr_{\mathcal{I}}(\mathrm{A}^\dag)(x,x')\mathrm{A}(x',x)
    = \tr_{\mathcal{I}}(\mathrm{A}^\dag\mathrm{A})(x,x) \\ &=
    \sum_{I,n} a_n^2 \phi_{n\,I}(x)\,\phi_{n\,I}(x)^* \le
    \left(\max_{n'} a_{n'}^2\right) \sum_{I,n}
    \phi_{n\,I}(x)\,\phi_{n\,I}(x)^* = \Vert \mathrm{A}\Vert^2
    \sum_{I}\delta_{II}\delta(x,x)\\ & = \mathrm{dim}\,
    \mathcal{I}\,\Vert \mathrm{A} \Vert^2 \,.
  \end{aligned}
\end{equation}

\subsection{Extension to absolutely integrable $\Mc_{\colA\colB}$}
\label{sec:abs_int}

The requirement of continuity of $\Mc_{\colA\colB}$, used in Section
\ref{sec:bounds} to bound $\Kc_\Mc$ independently of the gauge
configuration and uniformly in the volume, can be relaxed to the
requirement that the entries $\Mc_{\colA\colB}$ be absolutely
integrable, a condition conveniently expressed as
\begin{equation}
  \label{eq:L2_request}
  \left\la \Kc_\Mc(\colx,\coly)^{\f{1}{2}}\right\ra_{\textrm{G}}=\left\la 
    \left(\tr_{FCD}\{\Mc(\colx,\coly)
      \Mc(\colx,\coly)^\dag\}\right)^{\f{1}{2}}\right\ra_{\textrm{G}} <
  \infty\,. 
\end{equation}
Since $|X_1|\le \sqrt{\sum_i|X_i|^2}\le \sum_i|X_i|$, this condition
is equivalent to
$\left\la\left|\Mc_{\colA\colB}(\colx,\coly)\right|\right\ra_{\textrm{G}}
< \infty$ $\forall \colA,\colB$. It is further assumed that
Eq.~\eqref{eq:L2_request} remains true in the thermodynamic limit.

Under the assumption of a positive integration measure, and under
assumption (1.)\ in Section \ref{sec:not} on the norm of the
propagator, this suffices to prove convergence of
$\la\Oc_\Mc\ra$. Using this assumption in the first inequality in
Eq.~\eqref{eq:cs0_1} one finds
\begin{equation}
  \label{eq:cs0_1_app}
  \left|\left\la \Oc_\Mc    \right\ra\right| 
  \le n!\,\left(N_f N_c N_D\right)^{\f{n}{2}}
  \left\la
    \Kc_\Mc(\colx,\coly)^{\f{1}{2}}\Vert
    \propa_M\Vert^{n}\right\ra_{\mathrm{G}} 
  \le \f{n!\,\left(N_f N_c N_D\right)^{\f{n}{2}}}{m_0^n}
  \left\la
    \Kc_\Mc(\colx,\coly)^{\f{1}{2}}\right\ra_{\mathrm{G}}\,,
\end{equation}
which under the absolute-integrability condition discussed above is a
finite bound that remains so as $\Vc\to\infty$.

Using also assumption (2.)\ in Section \ref{sec:not} one can prove the
impossibility of vector flavor symmetry breaking. Starting from
Eq.~\eqref{eq:cs1}, and using Eqs.~\eqref{eq:bound_prop_good4} and
\eqref{eq:bound_comm_good3} and the assumptions on
$\Vert\propa_M \Vert$ and $\Vert \deltaD\Vert$, one finds
\begin{equation}
  \label{eq:bound_L2}
  \begin{aligned}
    \left| \f{\de}{\de\theta_a}\left\la \Oc_\Mc^{\bm{\theta}}
      \right\ra \Big|_{\bm{\theta}=0}\right| &\le \delta m\,
    \bar{\Csc}_n \left\la \Kc_\Mc(\colx,\coly)^{\f{1}{2}} \Vert
      \propa_M \Vert^{n+1} \Vert \deltaD \Vert\right\ra_{\mathrm{G}}
    \le \delta m \f{\bar{\Csc}_n\ded}{m_0^{n+1}} \left\la
      \Kc_\Mc(\colx,\coly)^{\f{1}{2}} \right\ra_{\mathrm{G}}\,,
  \end{aligned}
\end{equation}
where
$\bar{\Csc}_n\equiv n\, n!\left( N_f N_c N_D\right)^{\f{n}{2}}
\left(\f{1}{2N_f}\right)^{\f{1}{2}}$.  Since the last factor is finite
with a finite thermodynamic limit thanks to the assumption of absolute
integrability, the desired result, Eq.~\eqref{eq:bound_summary3},
follows.

The case of finite transformations is obtained by a straightforward
extension of the argument of Appendix \ref{sec:finflavtr}: since the
right-hand side of Eq.~\eqref{eq:bound_L2} is unchanged when replacing
$\Mc\to \Mc^{\alpha\bm{\theta}}$, one has
\begin{equation}
  \label{eq:bound_L2_app}
  \begin{aligned}
    \left|\left\la \Oc_\Mc^{\bm{\theta}} -\Oc_\Mc \right\ra \right| &
    \le \sum_a |\theta_a|\int_0^1 d\alpha\, \left|\left\la \Cc^a \,
        \Oc_{\Mc^{\alpha\bm{\theta}}} \right\ra\right| \le \delta m
    \f{\bar{\Csc}_n\ded}{m_0^{n+1}} \left\la
      \Kc_\Mc(\colx,\coly)^{\f{1}{2}} \right\ra_{\mathrm{G}} \sum_a
    |\theta_a|\,,
  \end{aligned}
\end{equation}
from which Eq.~\eqref{eq:wi1bis_app5} follows under the assumption of
absolute integrability.
\end{widetext}

\bibliographystyle{apsrev4-2}
\bibliography{references_vf_prd}

\begin{thebibliography}{80}%
\makeatletter
\providecommand \@ifxundefined [1]{%
 \@ifx{#1\undefined}
}%
\providecommand \@ifnum [1]{%
 \ifnum #1\expandafter \@firstoftwo
 \else \expandafter \@secondoftwo
 \fi
}%
\providecommand \@ifx [1]{%
 \ifx #1\expandafter \@firstoftwo
 \else \expandafter \@secondoftwo
 \fi
}%
\providecommand \natexlab [1]{#1}%
\providecommand \enquote  [1]{``#1''}%
\providecommand \bibnamefont  [1]{#1}%
\providecommand \bibfnamefont [1]{#1}%
\providecommand \citenamefont [1]{#1}%
\providecommand \href@noop [0]{\@secondoftwo}%
\providecommand \href [0]{\begingroup \@sanitize@url \@href}%
\providecommand \@href[1]{\@@startlink{#1}\@@href}%
\providecommand \@@href[1]{\endgroup#1\@@endlink}%
\providecommand \@sanitize@url [0]{\catcode `\\12\catcode `\$12\catcode
  `\&12\catcode `\#12\catcode `\^12\catcode `\_12\catcode `\%12\relax}%
\providecommand \@@startlink[1]{}%
\providecommand \@@endlink[0]{}%
\providecommand \url  [0]{\begingroup\@sanitize@url \@url }%
\providecommand \@url [1]{\endgroup\@href {#1}{\urlprefix }}%
\providecommand \urlprefix  [0]{URL }%
\providecommand \Eprint [0]{\href }%
\providecommand \doibase [0]{https://doi.org/}%
\providecommand \selectlanguage [0]{\@gobble}%
\providecommand \bibinfo  [0]{\@secondoftwo}%
\providecommand \bibfield  [0]{\@secondoftwo}%
\providecommand \translation [1]{[#1]}%
\providecommand \BibitemOpen [0]{}%
\providecommand \bibitemStop [0]{}%
\providecommand \bibitemNoStop [0]{.\EOS\space}%
\providecommand \EOS [0]{\spacefactor3000\relax}%
\providecommand \BibitemShut  [1]{\csname bibitem#1\endcsname}%
\let\auto@bib@innerbib\@empty
\bibitem [{\citenamefont {Vafa}\ and\ \citenamefont
  {Witten}(1984)}]{Vafa:1983tf}%
  \BibitemOpen
  \bibfield  {author} {\bibinfo {author} {\bibfnamefont {C.}~\bibnamefont
  {Vafa}}\ and\ \bibinfo {author} {\bibfnamefont {E.}~\bibnamefont {Witten}},\
  }\href {https://doi.org/10.1016/0550-3213(84)90230-X} {\bibfield  {journal}
  {\bibinfo  {journal} {Nucl. Phys. B}\ }\textbf {\bibinfo {volume} {234}},\
  \bibinfo {pages} {173} (\bibinfo {year} {1984})}\BibitemShut {NoStop}%
\bibitem [{\citenamefont {Goldstone}\ \emph {et~al.}(1962)\citenamefont
  {Goldstone}, \citenamefont {Salam},\ and\ \citenamefont
  {Weinberg}}]{Goldstone:1962es}%
  \BibitemOpen
  \bibfield  {author} {\bibinfo {author} {\bibfnamefont {J.}~\bibnamefont
  {Goldstone}}, \bibinfo {author} {\bibfnamefont {A.}~\bibnamefont {Salam}},\
  and\ \bibinfo {author} {\bibfnamefont {S.}~\bibnamefont {Weinberg}},\ }\href
  {https://doi.org/10.1103/PhysRev.127.965} {\bibfield  {journal} {\bibinfo
  {journal} {Phys. Rev.}\ }\textbf {\bibinfo {volume} {127}},\ \bibinfo {pages}
  {965} (\bibinfo {year} {1962})}\BibitemShut {NoStop}%
\bibitem [{\citenamefont {Lange}(1965)}]{Lange:1965zz}%
  \BibitemOpen
  \bibfield  {author} {\bibinfo {author} {\bibfnamefont {R.}~\bibnamefont
  {Lange}},\ }\href {https://doi.org/10.1103/PhysRevLett.14.3} {\bibfield
  {journal} {\bibinfo  {journal} {Phys. Rev. Lett.}\ }\textbf {\bibinfo
  {volume} {14}},\ \bibinfo {pages} {3} (\bibinfo {year} {1965})}\BibitemShut
  {NoStop}%
\bibitem [{\citenamefont {Strocchi}(2008)}]{Strocchi:2008gsa}%
  \BibitemOpen
  \bibfield  {author} {\bibinfo {author} {\bibfnamefont {F.}~\bibnamefont
  {Strocchi}},\ }\href {https://doi.org/10.1007/978-3-540-73593-9} {\emph
  {\bibinfo {title} {{Symmetry Breaking}}}},\ \bibinfo {series} {Lect. Notes
  Phys.}, Vol.\ \bibinfo {volume} {732}\ (\bibinfo  {publisher} {Springer},\
  \bibinfo {address} {Berlin},\ \bibinfo {year} {2008})\BibitemShut {NoStop}%
\bibitem [{\citenamefont {Aoki}(1984)}]{Aoki:1983qi}%
  \BibitemOpen
  \bibfield  {author} {\bibinfo {author} {\bibfnamefont {S.}~\bibnamefont
  {Aoki}},\ }\href {https://doi.org/10.1103/PhysRevD.30.2653} {\bibfield
  {journal} {\bibinfo  {journal} {Phys. Rev. D}\ }\textbf {\bibinfo {volume}
  {30}},\ \bibinfo {pages} {2653} (\bibinfo {year} {1984})}\BibitemShut
  {NoStop}%
\bibitem [{\citenamefont {Aoki}(1986)}]{Aoki:1986xr}%
  \BibitemOpen
  \bibfield  {author} {\bibinfo {author} {\bibfnamefont {S.}~\bibnamefont
  {Aoki}},\ }\href {https://doi.org/10.1103/PhysRevLett.57.3136} {\bibfield
  {journal} {\bibinfo  {journal} {Phys. Rev. Lett.}\ }\textbf {\bibinfo
  {volume} {57}},\ \bibinfo {pages} {3136} (\bibinfo {year}
  {1986})}\BibitemShut {NoStop}%
\bibitem [{\citenamefont {Setoodeh}\ \emph {et~al.}(1988)\citenamefont
  {Setoodeh}, \citenamefont {Davies},\ and\ \citenamefont
  {Barbour}}]{Setoodeh:1988ds}%
  \BibitemOpen
  \bibfield  {author} {\bibinfo {author} {\bibfnamefont {R.}~\bibnamefont
  {Setoodeh}}, \bibinfo {author} {\bibfnamefont {C.~T.~H.}\ \bibnamefont
  {Davies}},\ and\ \bibinfo {author} {\bibfnamefont {I.~M.}\ \bibnamefont
  {Barbour}},\ }\href {https://doi.org/10.1016/0370-2693(88)91025-8} {\bibfield
   {journal} {\bibinfo  {journal} {Phys. Lett. B}\ }\textbf {\bibinfo {volume}
  {213}},\ \bibinfo {pages} {195} (\bibinfo {year} {1988})}\BibitemShut
  {NoStop}%
\bibitem [{\citenamefont {Aoki}\ and\ \citenamefont
  {Gocksch}(1989)}]{Aoki:1989rw}%
  \BibitemOpen
  \bibfield  {author} {\bibinfo {author} {\bibfnamefont {S.}~\bibnamefont
  {Aoki}}\ and\ \bibinfo {author} {\bibfnamefont {A.}~\bibnamefont {Gocksch}},\
  }\href {https://doi.org/10.1016/0370-2693(89)90692-8} {\bibfield  {journal}
  {\bibinfo  {journal} {Phys. Lett. B}\ }\textbf {\bibinfo {volume} {231}},\
  \bibinfo {pages} {449} (\bibinfo {year} {1989})}\BibitemShut {NoStop}%
\bibitem [{\citenamefont {Aoki}\ and\ \citenamefont
  {Gocksch}(1990)}]{Aoki:1990ap}%
  \BibitemOpen
  \bibfield  {author} {\bibinfo {author} {\bibfnamefont {S.}~\bibnamefont
  {Aoki}}\ and\ \bibinfo {author} {\bibfnamefont {A.}~\bibnamefont {Gocksch}},\
  }\href {https://doi.org/10.1016/0370-2693(90)91405-Z} {\bibfield  {journal}
  {\bibinfo  {journal} {Phys. Lett. B}\ }\textbf {\bibinfo {volume} {243}},\
  \bibinfo {pages} {409} (\bibinfo {year} {1990})}\BibitemShut {NoStop}%
\bibitem [{\citenamefont {Aoki}\ and\ \citenamefont
  {Gocksch}(1992)}]{Aoki:1992nb}%
  \BibitemOpen
  \bibfield  {author} {\bibinfo {author} {\bibfnamefont {S.}~\bibnamefont
  {Aoki}}\ and\ \bibinfo {author} {\bibfnamefont {A.}~\bibnamefont {Gocksch}},\
  }\href {https://doi.org/10.1103/PhysRevD.45.3845} {\bibfield  {journal}
  {\bibinfo  {journal} {Phys. Rev. D}\ }\textbf {\bibinfo {volume} {45}},\
  \bibinfo {pages} {3845} (\bibinfo {year} {1992})}\BibitemShut {NoStop}%
\bibitem [{\citenamefont {Aoki}\ \emph
  {et~al.}(1996{\natexlab{a}})\citenamefont {Aoki}, \citenamefont {Ukawa},\
  and\ \citenamefont {Umemura}}]{Aoki:1995yf}%
  \BibitemOpen
  \bibfield  {author} {\bibinfo {author} {\bibfnamefont {S.}~\bibnamefont
  {Aoki}}, \bibinfo {author} {\bibfnamefont {A.}~\bibnamefont {Ukawa}},\ and\
  \bibinfo {author} {\bibfnamefont {T.}~\bibnamefont {Umemura}},\ }\href
  {https://doi.org/10.1103/PhysRevLett.76.873} {\bibfield  {journal} {\bibinfo
  {journal} {Phys. Rev. Lett.}\ }\textbf {\bibinfo {volume} {76}},\ \bibinfo
  {pages} {873} (\bibinfo {year} {1996}{\natexlab{a}})},\ \Eprint
  {https://arxiv.org/abs/hep-lat/9508008} {arXiv:hep-lat/9508008} \BibitemShut
  {NoStop}%
\bibitem [{\citenamefont {Aoki}(1996)}]{Aoki:1995ft}%
  \BibitemOpen
  \bibfield  {author} {\bibinfo {author} {\bibfnamefont {S.}~\bibnamefont
  {Aoki}},\ }\href {https://doi.org/10.1143/PTPS.122.179} {\bibfield  {journal}
  {\bibinfo  {journal} {Prog. Theor. Phys. Suppl.}\ }\textbf {\bibinfo {volume}
  {122}},\ \bibinfo {pages} {179} (\bibinfo {year} {1996})},\ \Eprint
  {https://arxiv.org/abs/hep-lat/9509008} {arXiv:hep-lat/9509008} \BibitemShut
  {NoStop}%
\bibitem [{\citenamefont {Aoki}\ \emph
  {et~al.}(1996{\natexlab{b}})\citenamefont {Aoki}, \citenamefont {Ukawa},\
  and\ \citenamefont {Umemura}}]{Aoki:1995ba}%
  \BibitemOpen
  \bibfield  {author} {\bibinfo {author} {\bibfnamefont {S.}~\bibnamefont
  {Aoki}}, \bibinfo {author} {\bibfnamefont {A.}~\bibnamefont {Ukawa}},\ and\
  \bibinfo {author} {\bibfnamefont {T.}~\bibnamefont {Umemura}},\ }\href
  {https://doi.org/10.1016/0920-5632(96)00111-9} {\bibfield  {journal}
  {\bibinfo  {journal} {Nucl. Phys. B Proc. Suppl.}\ }\textbf {\bibinfo
  {volume} {47}},\ \bibinfo {pages} {511} (\bibinfo {year}
  {1996}{\natexlab{b}})},\ \Eprint {https://arxiv.org/abs/hep-lat/9510014}
  {arXiv:hep-lat/9510014} \BibitemShut {NoStop}%
\bibitem [{\citenamefont {Aoki}\ \emph
  {et~al.}(1997{\natexlab{a}})\citenamefont {Aoki}, \citenamefont {Kaneda},
  \citenamefont {Ukawa},\ and\ \citenamefont {Umemura}}]{Aoki:1996pw}%
  \BibitemOpen
  \bibfield  {author} {\bibinfo {author} {\bibfnamefont {S.}~\bibnamefont
  {Aoki}}, \bibinfo {author} {\bibfnamefont {T.}~\bibnamefont {Kaneda}},
  \bibinfo {author} {\bibfnamefont {A.}~\bibnamefont {Ukawa}},\ and\ \bibinfo
  {author} {\bibfnamefont {T.}~\bibnamefont {Umemura}},\ }\href
  {https://doi.org/10.1016/S0920-5632(96)00682-2} {\bibfield  {journal}
  {\bibinfo  {journal} {Nucl. Phys. B Proc. Suppl.}\ }\textbf {\bibinfo
  {volume} {53}},\ \bibinfo {pages} {438} (\bibinfo {year}
  {1997}{\natexlab{a}})},\ \Eprint {https://arxiv.org/abs/hep-lat/9612010}
  {arXiv:hep-lat/9612010} \BibitemShut {NoStop}%
\bibitem [{\citenamefont {Aoki}\ \emph
  {et~al.}(1997{\natexlab{b}})\citenamefont {Aoki}, \citenamefont {Kaneda},\
  and\ \citenamefont {Ukawa}}]{Aoki:1996af}%
  \BibitemOpen
  \bibfield  {author} {\bibinfo {author} {\bibfnamefont {S.}~\bibnamefont
  {Aoki}}, \bibinfo {author} {\bibfnamefont {T.}~\bibnamefont {Kaneda}},\ and\
  \bibinfo {author} {\bibfnamefont {A.}~\bibnamefont {Ukawa}},\ }\href
  {https://doi.org/10.1103/PhysRevD.56.1808} {\bibfield  {journal} {\bibinfo
  {journal} {Phys. Rev. D}\ }\textbf {\bibinfo {volume} {56}},\ \bibinfo
  {pages} {1808} (\bibinfo {year} {1997}{\natexlab{b}})},\ \Eprint
  {https://arxiv.org/abs/hep-lat/9612019} {arXiv:hep-lat/9612019} \BibitemShut
  {NoStop}%
\bibitem [{\citenamefont {Aoki}(1998)}]{Aoki:1997fm}%
  \BibitemOpen
  \bibfield  {author} {\bibinfo {author} {\bibfnamefont {S.}~\bibnamefont
  {Aoki}},\ }\href {https://doi.org/10.1016/S0920-5632(97)00483-0} {\bibfield
  {journal} {\bibinfo  {journal} {Nucl. Phys. B Proc. Suppl.}\ }\textbf
  {\bibinfo {volume} {60}},\ \bibinfo {pages} {206} (\bibinfo {year} {1998})},\
  \Eprint {https://arxiv.org/abs/hep-lat/9707020} {arXiv:hep-lat/9707020}
  \BibitemShut {NoStop}%
\bibitem [{\citenamefont {Bitar}(1997)}]{Bitar:1996kc}%
  \BibitemOpen
  \bibfield  {author} {\bibinfo {author} {\bibfnamefont {K.~M.}\ \bibnamefont
  {Bitar}},\ }\href {https://doi.org/10.1103/PhysRevD.56.2736} {\bibfield
  {journal} {\bibinfo  {journal} {Phys. Rev. D}\ }\textbf {\bibinfo {volume}
  {56}},\ \bibinfo {pages} {2736} (\bibinfo {year} {1997})},\ \Eprint
  {https://arxiv.org/abs/hep-lat/9602027} {arXiv:hep-lat/9602027} \BibitemShut
  {NoStop}%
\bibitem [{\citenamefont {Bitar}(1998)}]{Bitar:1997as}%
  \BibitemOpen
  \bibfield  {author} {\bibinfo {author} {\bibfnamefont {K.~M.}\ \bibnamefont
  {Bitar}},\ }\href {https://doi.org/10.1016/S0920-5632(97)00913-4} {\bibfield
  {journal} {\bibinfo  {journal} {Nucl. Phys. B Proc. Suppl.}\ }\textbf
  {\bibinfo {volume} {63}},\ \bibinfo {pages} {829} (\bibinfo {year} {1998})},\
  \Eprint {https://arxiv.org/abs/hep-lat/9709086} {arXiv:hep-lat/9709086}
  \BibitemShut {NoStop}%
\bibitem [{\citenamefont {Bitar}\ \emph {et~al.}(1998)\citenamefont {Bitar},
  \citenamefont {Heller},\ and\ \citenamefont {Narayanan}}]{Bitar:1997ic}%
  \BibitemOpen
  \bibfield  {author} {\bibinfo {author} {\bibfnamefont {K.~M.}\ \bibnamefont
  {Bitar}}, \bibinfo {author} {\bibfnamefont {U.~M.}\ \bibnamefont {Heller}},\
  and\ \bibinfo {author} {\bibfnamefont {R.}~\bibnamefont {Narayanan}},\ }\href
  {https://doi.org/10.1016/S0370-2693(97)01487-1} {\bibfield  {journal}
  {\bibinfo  {journal} {Phys. Lett. B}\ }\textbf {\bibinfo {volume} {418}},\
  \bibinfo {pages} {167} (\bibinfo {year} {1998})},\ \Eprint
  {https://arxiv.org/abs/hep-th/9710052} {arXiv:hep-th/9710052} \BibitemShut
  {NoStop}%
\bibitem [{\citenamefont {Edwards}\ \emph
  {et~al.}(1998{\natexlab{a}})\citenamefont {Edwards}, \citenamefont {Heller},
  \citenamefont {Narayanan},\ and\ \citenamefont {Singleton}}]{Edwards:1997sp}%
  \BibitemOpen
  \bibfield  {author} {\bibinfo {author} {\bibfnamefont {R.~G.}\ \bibnamefont
  {Edwards}}, \bibinfo {author} {\bibfnamefont {U.~M.}\ \bibnamefont {Heller}},
  \bibinfo {author} {\bibfnamefont {R.}~\bibnamefont {Narayanan}},\ and\
  \bibinfo {author} {\bibfnamefont {R.~L.}\ \bibnamefont {Singleton},
  \bibfnamefont {Jr.}},\ }\href {https://doi.org/10.1016/S0550-3213(98)00104-7}
  {\bibfield  {journal} {\bibinfo  {journal} {Nucl. Phys. B}\ }\textbf
  {\bibinfo {volume} {518}},\ \bibinfo {pages} {319} (\bibinfo {year}
  {1998}{\natexlab{a}})},\ \Eprint {https://arxiv.org/abs/hep-lat/9711029}
  {arXiv:hep-lat/9711029} \BibitemShut {NoStop}%
\bibitem [{\citenamefont {Edwards}\ \emph
  {et~al.}(1998{\natexlab{b}})\citenamefont {Edwards}, \citenamefont {Heller},\
  and\ \citenamefont {Narayanan}}]{Edwards:1998sh}%
  \BibitemOpen
  \bibfield  {author} {\bibinfo {author} {\bibfnamefont {R.~G.}\ \bibnamefont
  {Edwards}}, \bibinfo {author} {\bibfnamefont {U.~M.}\ \bibnamefont
  {Heller}},\ and\ \bibinfo {author} {\bibfnamefont {R.}~\bibnamefont
  {Narayanan}},\ }\href {https://doi.org/10.1016/S0550-3213(98)00588-4}
  {\bibfield  {journal} {\bibinfo  {journal} {Nucl. Phys. B}\ }\textbf
  {\bibinfo {volume} {535}},\ \bibinfo {pages} {403} (\bibinfo {year}
  {1998}{\natexlab{b}})},\ \Eprint {https://arxiv.org/abs/hep-lat/9802016}
  {arXiv:hep-lat/9802016} \BibitemShut {NoStop}%
\bibitem [{\citenamefont {Sharpe}\ and\ \citenamefont
  {Singleton}(1998)}]{Sharpe:1998xm}%
  \BibitemOpen
  \bibfield  {author} {\bibinfo {author} {\bibfnamefont {S.~R.}\ \bibnamefont
  {Sharpe}}\ and\ \bibinfo {author} {\bibfnamefont {R.~L.}\ \bibnamefont
  {Singleton}, \bibfnamefont {Jr.}},\ }\href
  {https://doi.org/10.1103/PhysRevD.58.074501} {\bibfield  {journal} {\bibinfo
  {journal} {Phys. Rev. D}\ }\textbf {\bibinfo {volume} {58}},\ \bibinfo
  {pages} {074501} (\bibinfo {year} {1998})},\ \Eprint
  {https://arxiv.org/abs/hep-lat/9804028} {arXiv:hep-lat/9804028} \BibitemShut
  {NoStop}%
\bibitem [{\citenamefont {Azcoiti}\ \emph {et~al.}(2009)\citenamefont
  {Azcoiti}, \citenamefont {Di~Carlo},\ and\ \citenamefont
  {Vaquero}}]{Azcoiti:2008dn}%
  \BibitemOpen
  \bibfield  {author} {\bibinfo {author} {\bibfnamefont {V.}~\bibnamefont
  {Azcoiti}}, \bibinfo {author} {\bibfnamefont {G.}~\bibnamefont {Di~Carlo}},\
  and\ \bibinfo {author} {\bibfnamefont {A.}~\bibnamefont {Vaquero}},\ }\href
  {https://doi.org/10.1103/PhysRevD.79.014509} {\bibfield  {journal} {\bibinfo
  {journal} {Phys. Rev. D}\ }\textbf {\bibinfo {volume} {79}},\ \bibinfo
  {pages} {014509} (\bibinfo {year} {2009})},\ \Eprint
  {https://arxiv.org/abs/0809.2972} {arXiv:0809.2972 [hep-lat]} \BibitemShut
  {NoStop}%
\bibitem [{\citenamefont {Sharpe}(2009)}]{Sharpe:2008ke}%
  \BibitemOpen
  \bibfield  {author} {\bibinfo {author} {\bibfnamefont {S.~R.}\ \bibnamefont
  {Sharpe}},\ }\href {https://doi.org/10.1103/PhysRevD.79.054503} {\bibfield
  {journal} {\bibinfo  {journal} {Phys. Rev. D}\ }\textbf {\bibinfo {volume}
  {79}},\ \bibinfo {pages} {054503} (\bibinfo {year} {2009})},\ \Eprint
  {https://arxiv.org/abs/0811.0409} {arXiv:0811.0409 [hep-lat]} \BibitemShut
  {NoStop}%
\bibitem [{\citenamefont {Azcoiti}\ \emph {et~al.}(2013)\citenamefont
  {Azcoiti}, \citenamefont {Di~Carlo}, \citenamefont {Follana},\ and\
  \citenamefont {Vaquero}}]{Azcoiti:2012ns}%
  \BibitemOpen
  \bibfield  {author} {\bibinfo {author} {\bibfnamefont {V.}~\bibnamefont
  {Azcoiti}}, \bibinfo {author} {\bibfnamefont {G.}~\bibnamefont {Di~Carlo}},
  \bibinfo {author} {\bibfnamefont {E.}~\bibnamefont {Follana}},\ and\ \bibinfo
  {author} {\bibfnamefont {A.}~\bibnamefont {Vaquero}},\ }\href
  {https://doi.org/10.1016/j.nuclphysb.2013.01.008} {\bibfield  {journal}
  {\bibinfo  {journal} {Nucl. Phys. B}\ }\textbf {\bibinfo {volume} {870}},\
  \bibinfo {pages} {138} (\bibinfo {year} {2013})},\ \Eprint
  {https://arxiv.org/abs/1208.0761} {arXiv:1208.0761 [hep-lat]} \BibitemShut
  {NoStop}%
\bibitem [{\citenamefont {Wilson}(1977)}]{Wilson1977}%
  \BibitemOpen
  \bibfield  {author} {\bibinfo {author} {\bibfnamefont {K.~G.}\ \bibnamefont
  {Wilson}},\ }in\ \href {https://doi.org/10.1007/978-1-4613-4208-3_6} {\emph
  {\bibinfo {booktitle} {New Phenomena in Subnuclear Physics: Part A}}},\
  \bibinfo {editor} {edited by\ \bibinfo {editor} {\bibfnamefont
  {A.}~\bibnamefont {Zichichi}}}\ (\bibinfo  {publisher} {Springer US},\
  \bibinfo {address} {Boston, MA},\ \bibinfo {year} {1977})\ pp.\ \bibinfo
  {pages} {69--142}\BibitemShut {NoStop}%
\bibitem [{\citenamefont {Kogut}\ and\ \citenamefont
  {Susskind}(1975)}]{Kogut:1974ag}%
  \BibitemOpen
  \bibfield  {author} {\bibinfo {author} {\bibfnamefont {J.~B.}\ \bibnamefont
  {Kogut}}\ and\ \bibinfo {author} {\bibfnamefont {L.}~\bibnamefont
  {Susskind}},\ }\href {https://doi.org/10.1103/PhysRevD.11.395} {\bibfield
  {journal} {\bibinfo  {journal} {Phys. Rev. D}\ }\textbf {\bibinfo {volume}
  {11}},\ \bibinfo {pages} {395} (\bibinfo {year} {1975})}\BibitemShut
  {NoStop}%
\bibitem [{\citenamefont {Susskind}(1977)}]{Susskind:1976jm}%
  \BibitemOpen
  \bibfield  {author} {\bibinfo {author} {\bibfnamefont {L.}~\bibnamefont
  {Susskind}},\ }\href {https://doi.org/10.1103/PhysRevD.16.3031} {\bibfield
  {journal} {\bibinfo  {journal} {Phys. Rev. D}\ }\textbf {\bibinfo {volume}
  {16}},\ \bibinfo {pages} {3031} (\bibinfo {year} {1977})}\BibitemShut
  {NoStop}%
\bibitem [{\citenamefont {Banks}\ \emph {et~al.}(1977)\citenamefont {Banks},
  \citenamefont {Raby}, \citenamefont {Susskind}, \citenamefont {Kogut},
  \citenamefont {Jones}, \citenamefont {Scharbach},\ and\ \citenamefont
  {Sinclair}}]{Banks:1976ia}%
  \BibitemOpen
  \bibfield  {author} {\bibinfo {author} {\bibfnamefont {T.}~\bibnamefont
  {Banks}}, \bibinfo {author} {\bibfnamefont {S.}~\bibnamefont {Raby}},
  \bibinfo {author} {\bibfnamefont {L.}~\bibnamefont {Susskind}}, \bibinfo
  {author} {\bibfnamefont {J.~B.}\ \bibnamefont {Kogut}}, \bibinfo {author}
  {\bibfnamefont {D.~R.~T.}\ \bibnamefont {Jones}}, \bibinfo {author}
  {\bibfnamefont {P.~N.}\ \bibnamefont {Scharbach}},\ and\ \bibinfo {author}
  {\bibfnamefont {D.~K.}\ \bibnamefont {Sinclair}} (\bibinfo {collaboration}
  {Cornell-Oxford-Tel Aviv-Yeshiva Collaboration}),\ }\href
  {https://doi.org/10.1103/PhysRevD.15.1111} {\bibfield  {journal} {\bibinfo
  {journal} {Phys. Rev. D}\ }\textbf {\bibinfo {volume} {15}},\ \bibinfo
  {pages} {1111} (\bibinfo {year} {1977})}\BibitemShut {NoStop}%
\bibitem [{\citenamefont {Aloisio}\ \emph {et~al.}(2001)\citenamefont
  {Aloisio}, \citenamefont {Azcoiti}, \citenamefont {Di~Carlo}, \citenamefont
  {Galante},\ and\ \citenamefont {Grillo}}]{Aloisio:2000rb}%
  \BibitemOpen
  \bibfield  {author} {\bibinfo {author} {\bibfnamefont {R.}~\bibnamefont
  {Aloisio}}, \bibinfo {author} {\bibfnamefont {V.}~\bibnamefont {Azcoiti}},
  \bibinfo {author} {\bibfnamefont {G.}~\bibnamefont {Di~Carlo}}, \bibinfo
  {author} {\bibfnamefont {A.}~\bibnamefont {Galante}},\ and\ \bibinfo {author}
  {\bibfnamefont {A.~F.}\ \bibnamefont {Grillo}},\ }\href
  {https://doi.org/10.1016/S0550-3213(01)00232-2} {\bibfield  {journal}
  {\bibinfo  {journal} {Nucl. Phys. B}\ }\textbf {\bibinfo {volume} {606}},\
  \bibinfo {pages} {322} (\bibinfo {year} {2001})},\ \Eprint
  {https://arxiv.org/abs/hep-lat/0011079} {arXiv:hep-lat/0011079} \BibitemShut
  {NoStop}%
\bibitem [{\citenamefont {Hamber}\ \emph {et~al.}(1983)\citenamefont {Hamber},
  \citenamefont {Marinari}, \citenamefont {Parisi},\ and\ \citenamefont
  {Rebbi}}]{Hamber:1983kx}%
  \BibitemOpen
  \bibfield  {author} {\bibinfo {author} {\bibfnamefont {H.~W.}\ \bibnamefont
  {Hamber}}, \bibinfo {author} {\bibfnamefont {E.}~\bibnamefont {Marinari}},
  \bibinfo {author} {\bibfnamefont {G.}~\bibnamefont {Parisi}},\ and\ \bibinfo
  {author} {\bibfnamefont {C.}~\bibnamefont {Rebbi}},\ }\href
  {https://doi.org/10.1016/0370-2693(83)91412-0} {\bibfield  {journal}
  {\bibinfo  {journal} {Phys. Lett. B}\ }\textbf {\bibinfo {volume} {124}},\
  \bibinfo {pages} {99} (\bibinfo {year} {1983})}\BibitemShut {NoStop}%
\bibitem [{\citenamefont {Fucito}\ and\ \citenamefont
  {Solomon}(1984)}]{Fucito:1984nu}%
  \BibitemOpen
  \bibfield  {author} {\bibinfo {author} {\bibfnamefont {F.}~\bibnamefont
  {Fucito}}\ and\ \bibinfo {author} {\bibfnamefont {S.}~\bibnamefont
  {Solomon}},\ }\href {https://doi.org/10.1016/0370-2693(84)90777-9} {\bibfield
   {journal} {\bibinfo  {journal} {Phys. Lett. B}\ }\textbf {\bibinfo {volume}
  {140}},\ \bibinfo {pages} {387} (\bibinfo {year} {1984})}\BibitemShut
  {NoStop}%
\bibitem [{\citenamefont {Gottlieb}\ \emph {et~al.}(1988)\citenamefont
  {Gottlieb}, \citenamefont {Liu}, \citenamefont {Renken}, \citenamefont
  {Sugar},\ and\ \citenamefont {Toussaint}}]{Gottlieb:1988gr}%
  \BibitemOpen
  \bibfield  {author} {\bibinfo {author} {\bibfnamefont {S.~A.}\ \bibnamefont
  {Gottlieb}}, \bibinfo {author} {\bibfnamefont {W.}~\bibnamefont {Liu}},
  \bibinfo {author} {\bibfnamefont {R.~L.}\ \bibnamefont {Renken}}, \bibinfo
  {author} {\bibfnamefont {R.~L.}\ \bibnamefont {Sugar}},\ and\ \bibinfo
  {author} {\bibfnamefont {D.}~\bibnamefont {Toussaint}},\ }\href
  {https://doi.org/10.1103/PhysRevD.38.2245} {\bibfield  {journal} {\bibinfo
  {journal} {Phys. Rev. D}\ }\textbf {\bibinfo {volume} {38}},\ \bibinfo
  {pages} {2245} (\bibinfo {year} {1988})}\BibitemShut {NoStop}%
\bibitem [{\citenamefont {Creutz}(2006{\natexlab{a}})}]{Creutz:2006ys}%
  \BibitemOpen
  \bibfield  {author} {\bibinfo {author} {\bibfnamefont {M.}~\bibnamefont
  {Creutz}},\ }\Eprint {https://arxiv.org/abs/hep-lat/0603020}
  {arXiv:hep-lat/0603020}  (\bibinfo {year} {2006}{\natexlab{a}})\BibitemShut
  {NoStop}%
\bibitem [{\citenamefont {Creutz}(2006{\natexlab{b}})}]{Creutz:2006wv}%
  \BibitemOpen
  \bibfield  {author} {\bibinfo {author} {\bibfnamefont {M.}~\bibnamefont
  {Creutz}},\ }\href {https://doi.org/10.22323/1.032.0208} {\bibfield
  {journal} {\bibinfo  {journal} {PoS}\ }\textbf {\bibinfo {volume}
  {LAT2006}},\ \bibinfo {pages} {208} (\bibinfo {year} {2006}{\natexlab{b}})},\
  \Eprint {https://arxiv.org/abs/hep-lat/0608020} {arXiv:hep-lat/0608020}
  \BibitemShut {NoStop}%
\bibitem [{\citenamefont {Creutz}(2007{\natexlab{a}})}]{Creutz:2007yg}%
  \BibitemOpen
  \bibfield  {author} {\bibinfo {author} {\bibfnamefont {M.}~\bibnamefont
  {Creutz}},\ }\href {https://doi.org/10.1016/j.physletb.2007.03.065}
  {\bibfield  {journal} {\bibinfo  {journal} {Phys. Lett. B}\ }\textbf
  {\bibinfo {volume} {649}},\ \bibinfo {pages} {230} (\bibinfo {year}
  {2007}{\natexlab{a}})},\ \Eprint {https://arxiv.org/abs/hep-lat/0701018}
  {arXiv:hep-lat/0701018} \BibitemShut {NoStop}%
\bibitem [{\citenamefont {Creutz}(2007{\natexlab{b}})}]{Creutz:2007pr}%
  \BibitemOpen
  \bibfield  {author} {\bibinfo {author} {\bibfnamefont {M.}~\bibnamefont
  {Creutz}},\ }\href {https://doi.org/10.1016/j.physletb.2007.04.017}
  {\bibfield  {journal} {\bibinfo  {journal} {Phys. Lett. B}\ }\textbf
  {\bibinfo {volume} {649}},\ \bibinfo {pages} {241} (\bibinfo {year}
  {2007}{\natexlab{b}})},\ \Eprint {https://arxiv.org/abs/0704.2016}
  {arXiv:0704.2016 [hep-lat]} \BibitemShut {NoStop}%
\bibitem [{\citenamefont {Creutz}(2007{\natexlab{c}})}]{Creutz:2007rk}%
  \BibitemOpen
  \bibfield  {author} {\bibinfo {author} {\bibfnamefont {M.}~\bibnamefont
  {Creutz}},\ }\href {https://doi.org/10.22323/1.042.0007} {\bibfield
  {journal} {\bibinfo  {journal} {PoS}\ }\textbf {\bibinfo {volume}
  {LATTICE2007}},\ \bibinfo {pages} {007} (\bibinfo {year}
  {2007}{\natexlab{c}})},\ \Eprint {https://arxiv.org/abs/0708.1295}
  {arXiv:0708.1295 [hep-lat]} \BibitemShut {NoStop}%
\bibitem [{\citenamefont {Creutz}(2008{\natexlab{a}})}]{Creutz:2008kb}%
  \BibitemOpen
  \bibfield  {author} {\bibinfo {author} {\bibfnamefont {M.}~\bibnamefont
  {Creutz}},\ }\href {https://doi.org/10.1103/PhysRevD.78.078501} {\bibfield
  {journal} {\bibinfo  {journal} {Phys. Rev. D}\ }\textbf {\bibinfo {volume}
  {78}},\ \bibinfo {pages} {078501} (\bibinfo {year} {2008}{\natexlab{a}})},\
  \Eprint {https://arxiv.org/abs/0805.1350} {arXiv:0805.1350 [hep-lat]}
  \BibitemShut {NoStop}%
\bibitem [{\citenamefont {Shamir}(2005)}]{Shamir:2004zc}%
  \BibitemOpen
  \bibfield  {author} {\bibinfo {author} {\bibfnamefont {Y.}~\bibnamefont
  {Shamir}},\ }\href {https://doi.org/10.1103/PhysRevD.71.034509} {\bibfield
  {journal} {\bibinfo  {journal} {Phys. Rev. D}\ }\textbf {\bibinfo {volume}
  {71}},\ \bibinfo {pages} {034509} (\bibinfo {year} {2005})},\ \Eprint
  {https://arxiv.org/abs/hep-lat/0412014} {arXiv:hep-lat/0412014} \BibitemShut
  {NoStop}%
\bibitem [{\citenamefont {Bernard}(2006)}]{Bernard:2006zw}%
  \BibitemOpen
  \bibfield  {author} {\bibinfo {author} {\bibfnamefont {C.}~\bibnamefont
  {Bernard}},\ }\href {https://doi.org/10.1103/PhysRevD.73.114503} {\bibfield
  {journal} {\bibinfo  {journal} {Phys. Rev. D}\ }\textbf {\bibinfo {volume}
  {73}},\ \bibinfo {pages} {114503} (\bibinfo {year} {2006})},\ \Eprint
  {https://arxiv.org/abs/hep-lat/0603011} {arXiv:hep-lat/0603011} \BibitemShut
  {NoStop}%
\bibitem [{\citenamefont {Bernard}\ \emph {et~al.}(2007)\citenamefont
  {Bernard}, \citenamefont {Golterman}, \citenamefont {Shamir},\ and\
  \citenamefont {Sharpe}}]{Bernard:2006vv}%
  \BibitemOpen
  \bibfield  {author} {\bibinfo {author} {\bibfnamefont {C.}~\bibnamefont
  {Bernard}}, \bibinfo {author} {\bibfnamefont {M.}~\bibnamefont {Golterman}},
  \bibinfo {author} {\bibfnamefont {Y.}~\bibnamefont {Shamir}},\ and\ \bibinfo
  {author} {\bibfnamefont {S.~R.}\ \bibnamefont {Sharpe}},\ }\href
  {https://doi.org/10.1016/j.physletb.2007.04.018} {\bibfield  {journal}
  {\bibinfo  {journal} {Phys. Lett. B}\ }\textbf {\bibinfo {volume} {649}},\
  \bibinfo {pages} {235} (\bibinfo {year} {2007})},\ \Eprint
  {https://arxiv.org/abs/hep-lat/0603027} {arXiv:hep-lat/0603027} \BibitemShut
  {NoStop}%
\bibitem [{\citenamefont {Shamir}(2007)}]{Shamir:2006nj}%
  \BibitemOpen
  \bibfield  {author} {\bibinfo {author} {\bibfnamefont {Y.}~\bibnamefont
  {Shamir}},\ }\href {https://doi.org/10.1103/PhysRevD.75.054503} {\bibfield
  {journal} {\bibinfo  {journal} {Phys. Rev. D}\ }\textbf {\bibinfo {volume}
  {75}},\ \bibinfo {pages} {054503} (\bibinfo {year} {2007})},\ \Eprint
  {https://arxiv.org/abs/hep-lat/0607007} {arXiv:hep-lat/0607007} \BibitemShut
  {NoStop}%
\bibitem [{\citenamefont {Bernard}\ \emph
  {et~al.}(2008{\natexlab{a}})\citenamefont {Bernard}, \citenamefont
  {Golterman}, \citenamefont {Shamir},\ and\ \citenamefont
  {Sharpe}}]{Bernard:2007eh}%
  \BibitemOpen
  \bibfield  {author} {\bibinfo {author} {\bibfnamefont {C.}~\bibnamefont
  {Bernard}}, \bibinfo {author} {\bibfnamefont {M.}~\bibnamefont {Golterman}},
  \bibinfo {author} {\bibfnamefont {Y.}~\bibnamefont {Shamir}},\ and\ \bibinfo
  {author} {\bibfnamefont {S.~R.}\ \bibnamefont {Sharpe}},\ }\href
  {https://doi.org/10.1103/PhysRevD.77.114504} {\bibfield  {journal} {\bibinfo
  {journal} {Phys. Rev. D}\ }\textbf {\bibinfo {volume} {77}},\ \bibinfo
  {pages} {114504} (\bibinfo {year} {2008}{\natexlab{a}})},\ \Eprint
  {https://arxiv.org/abs/0711.0696} {arXiv:0711.0696 [hep-lat]} \BibitemShut
  {NoStop}%
\bibitem [{\citenamefont {Adams}(2008)}]{Adams:2008db}%
  \BibitemOpen
  \bibfield  {author} {\bibinfo {author} {\bibfnamefont {D.~H.}\ \bibnamefont
  {Adams}},\ }\href {https://doi.org/10.1103/PhysRevD.77.105024} {\bibfield
  {journal} {\bibinfo  {journal} {Phys. Rev. D}\ }\textbf {\bibinfo {volume}
  {77}},\ \bibinfo {pages} {105024} (\bibinfo {year} {2008})},\ \Eprint
  {https://arxiv.org/abs/0802.3029} {arXiv:0802.3029 [hep-lat]} \BibitemShut
  {NoStop}%
\bibitem [{\citenamefont {Bernard}\ \emph
  {et~al.}(2008{\natexlab{b}})\citenamefont {Bernard}, \citenamefont
  {Golterman}, \citenamefont {Shamir},\ and\ \citenamefont
  {Sharpe}}]{Bernard:2008gr}%
  \BibitemOpen
  \bibfield  {author} {\bibinfo {author} {\bibfnamefont {C.}~\bibnamefont
  {Bernard}}, \bibinfo {author} {\bibfnamefont {M.}~\bibnamefont {Golterman}},
  \bibinfo {author} {\bibfnamefont {Y.}~\bibnamefont {Shamir}},\ and\ \bibinfo
  {author} {\bibfnamefont {S.~R.}\ \bibnamefont {Sharpe}},\ }\href
  {https://doi.org/10.1103/PhysRevD.78.078502} {\bibfield  {journal} {\bibinfo
  {journal} {Phys. Rev. D}\ }\textbf {\bibinfo {volume} {78}},\ \bibinfo
  {pages} {078502} (\bibinfo {year} {2008}{\natexlab{b}})},\ \Eprint
  {https://arxiv.org/abs/0808.2056} {arXiv:0808.2056 [hep-lat]} \BibitemShut
  {NoStop}%
\bibitem [{\citenamefont {D{\"u}rr}(2006)}]{Durr:2005ax}%
  \BibitemOpen
  \bibfield  {author} {\bibinfo {author} {\bibfnamefont {S.}~\bibnamefont
  {D{\"u}rr}},\ }\href {https://doi.org/10.22323/1.020.0021} {\bibfield
  {journal} {\bibinfo  {journal} {PoS}\ }\textbf {\bibinfo {volume}
  {LAT2005}},\ \bibinfo {pages} {021} (\bibinfo {year} {2006})},\ \Eprint
  {https://arxiv.org/abs/hep-lat/0509026} {arXiv:hep-lat/0509026} \BibitemShut
  {NoStop}%
\bibitem [{\citenamefont {Sharpe}(2006)}]{Sharpe:2006re}%
  \BibitemOpen
  \bibfield  {author} {\bibinfo {author} {\bibfnamefont {S.~R.}\ \bibnamefont
  {Sharpe}},\ }\href {https://doi.org/10.22323/1.032.0022} {\bibfield
  {journal} {\bibinfo  {journal} {PoS}\ }\textbf {\bibinfo {volume}
  {LAT2006}},\ \bibinfo {pages} {022} (\bibinfo {year} {2006})},\ \Eprint
  {https://arxiv.org/abs/hep-lat/0610094} {arXiv:hep-lat/0610094} \BibitemShut
  {NoStop}%
\bibitem [{\citenamefont {Bernard}\ \emph {et~al.}(2006)\citenamefont
  {Bernard}, \citenamefont {Golterman},\ and\ \citenamefont
  {Shamir}}]{Bernard:2006qt}%
  \BibitemOpen
  \bibfield  {author} {\bibinfo {author} {\bibfnamefont {C.}~\bibnamefont
  {Bernard}}, \bibinfo {author} {\bibfnamefont {M.}~\bibnamefont {Golterman}},\
  and\ \bibinfo {author} {\bibfnamefont {Y.}~\bibnamefont {Shamir}},\ }\href
  {https://doi.org/10.22323/1.032.0205} {\bibfield  {journal} {\bibinfo
  {journal} {PoS}\ }\textbf {\bibinfo {volume} {LAT2006}},\ \bibinfo {pages}
  {205} (\bibinfo {year} {2006})},\ \Eprint
  {https://arxiv.org/abs/hep-lat/0610003} {arXiv:hep-lat/0610003} \BibitemShut
  {NoStop}%
\bibitem [{\citenamefont {Kronfeld}(2007)}]{Kronfeld:2007ek}%
  \BibitemOpen
  \bibfield  {author} {\bibinfo {author} {\bibfnamefont {A.~S.}\ \bibnamefont
  {Kronfeld}},\ }\href {https://doi.org/10.22323/1.042.0016} {\bibfield
  {journal} {\bibinfo  {journal} {PoS}\ }\textbf {\bibinfo {volume}
  {LATTICE2007}},\ \bibinfo {pages} {016} (\bibinfo {year} {2007})},\ \Eprint
  {https://arxiv.org/abs/0711.0699} {arXiv:0711.0699 [hep-lat]} \BibitemShut
  {NoStop}%
\bibitem [{\citenamefont {Golterman}(2008)}]{Golterman:2008gt}%
  \BibitemOpen
  \bibfield  {author} {\bibinfo {author} {\bibfnamefont {M.}~\bibnamefont
  {Golterman}},\ }\href {https://doi.org/10.22323/1.077.0014} {\bibfield
  {journal} {\bibinfo  {journal} {PoS}\ }\textbf {\bibinfo {volume}
  {CONFINEMENT8}},\ \bibinfo {pages} {014} (\bibinfo {year} {2008})},\ \Eprint
  {https://arxiv.org/abs/0812.3110} {arXiv:0812.3110 [hep-ph]} \BibitemShut
  {NoStop}%
\bibitem [{\citenamefont {Karsten}(1981)}]{Karsten:1981gd}%
  \BibitemOpen
  \bibfield  {author} {\bibinfo {author} {\bibfnamefont {L.~H.}\ \bibnamefont
  {Karsten}},\ }\href {https://doi.org/10.1016/0370-2693(81)90133-7} {\bibfield
   {journal} {\bibinfo  {journal} {Phys. Lett. B}\ }\textbf {\bibinfo {volume}
  {104}},\ \bibinfo {pages} {315} (\bibinfo {year} {1981})}\BibitemShut
  {NoStop}%
\bibitem [{\citenamefont {Wilczek}(1987)}]{Wilczek:1987kw}%
  \BibitemOpen
  \bibfield  {author} {\bibinfo {author} {\bibfnamefont {F.}~\bibnamefont
  {Wilczek}},\ }\href {https://doi.org/10.1103/PhysRevLett.59.2397} {\bibfield
  {journal} {\bibinfo  {journal} {Phys. Rev. Lett.}\ }\textbf {\bibinfo
  {volume} {59}},\ \bibinfo {pages} {2397} (\bibinfo {year}
  {1987})}\BibitemShut {NoStop}%
\bibitem [{\citenamefont {Creutz}(2008{\natexlab{b}})}]{Creutz:2007af}%
  \BibitemOpen
  \bibfield  {author} {\bibinfo {author} {\bibfnamefont {M.}~\bibnamefont
  {Creutz}},\ }\href {https://doi.org/10.1088/1126-6708/2008/04/017} {\bibfield
   {journal} {\bibinfo  {journal} {J. High Energy Phys.}\ }\textbf {\bibinfo
  {volume} {04}},\ \bibinfo {pages} {017}},\ \Eprint
  {https://arxiv.org/abs/0712.1201} {arXiv:0712.1201 [hep-lat]} \BibitemShut
  {NoStop}%
\bibitem [{\citenamefont {Bori{\c c}i}(2008)}]{Borici:2007kz}%
  \BibitemOpen
  \bibfield  {author} {\bibinfo {author} {\bibfnamefont {A.}~\bibnamefont
  {Bori{\c c}i}},\ }\href {https://doi.org/10.1103/PhysRevD.78.074504}
  {\bibfield  {journal} {\bibinfo  {journal} {Phys. Rev. D}\ }\textbf {\bibinfo
  {volume} {78}},\ \bibinfo {pages} {074504} (\bibinfo {year} {2008})},\
  \Eprint {https://arxiv.org/abs/0712.4401} {arXiv:0712.4401 [hep-lat]}
  \BibitemShut {NoStop}%
\bibitem [{\citenamefont {Ginsparg}\ and\ \citenamefont
  {Wilson}(1982)}]{Ginsparg:1981bj}%
  \BibitemOpen
  \bibfield  {author} {\bibinfo {author} {\bibfnamefont {P.~H.}\ \bibnamefont
  {Ginsparg}}\ and\ \bibinfo {author} {\bibfnamefont {K.~G.}\ \bibnamefont
  {Wilson}},\ }\href {https://doi.org/10.1103/PhysRevD.25.2649} {\bibfield
  {journal} {\bibinfo  {journal} {Phys. Rev. D}\ }\textbf {\bibinfo {volume}
  {25}},\ \bibinfo {pages} {2649} (\bibinfo {year} {1982})}\BibitemShut
  {NoStop}%
\bibitem [{\citenamefont {Hasenfratz}\ and\ \citenamefont
  {Niedermayer}(1994)}]{Hasenfratz:1993sp}%
  \BibitemOpen
  \bibfield  {author} {\bibinfo {author} {\bibfnamefont {P.}~\bibnamefont
  {Hasenfratz}}\ and\ \bibinfo {author} {\bibfnamefont {F.}~\bibnamefont
  {Niedermayer}},\ }\href {https://doi.org/10.1016/0550-3213(94)90261-5}
  {\bibfield  {journal} {\bibinfo  {journal} {Nucl. Phys. B}\ }\textbf
  {\bibinfo {volume} {414}},\ \bibinfo {pages} {785} (\bibinfo {year}
  {1994})},\ \Eprint {https://arxiv.org/abs/hep-lat/9308004}
  {arXiv:hep-lat/9308004} \BibitemShut {NoStop}%
\bibitem [{\citenamefont {DeGrand}\ \emph {et~al.}(1995)\citenamefont
  {DeGrand}, \citenamefont {Hasenfratz}, \citenamefont {Hasenfratz},\ and\
  \citenamefont {Niedermayer}}]{DeGrand:1995ji}%
  \BibitemOpen
  \bibfield  {author} {\bibinfo {author} {\bibfnamefont {T.~A.}\ \bibnamefont
  {DeGrand}}, \bibinfo {author} {\bibfnamefont {A.}~\bibnamefont {Hasenfratz}},
  \bibinfo {author} {\bibfnamefont {P.}~\bibnamefont {Hasenfratz}},\ and\
  \bibinfo {author} {\bibfnamefont {F.}~\bibnamefont {Niedermayer}},\ }\href
  {https://doi.org/10.1016/0550-3213(95)00458-5} {\bibfield  {journal}
  {\bibinfo  {journal} {Nucl. Phys. B}\ }\textbf {\bibinfo {volume} {454}},\
  \bibinfo {pages} {587} (\bibinfo {year} {1995})},\ \Eprint
  {https://arxiv.org/abs/hep-lat/9506030} {arXiv:hep-lat/9506030} \BibitemShut
  {NoStop}%
\bibitem [{\citenamefont {Hasenfratz}\ \emph {et~al.}(1998)\citenamefont
  {Hasenfratz}, \citenamefont {Laliena},\ and\ \citenamefont
  {Niedermayer}}]{Hasenfratz:1998ri}%
  \BibitemOpen
  \bibfield  {author} {\bibinfo {author} {\bibfnamefont {P.}~\bibnamefont
  {Hasenfratz}}, \bibinfo {author} {\bibfnamefont {V.}~\bibnamefont
  {Laliena}},\ and\ \bibinfo {author} {\bibfnamefont {F.}~\bibnamefont
  {Niedermayer}},\ }\href {https://doi.org/10.1016/S0370-2693(98)00315-3}
  {\bibfield  {journal} {\bibinfo  {journal} {Phys. Lett. B}\ }\textbf
  {\bibinfo {volume} {427}},\ \bibinfo {pages} {125} (\bibinfo {year}
  {1998})},\ \Eprint {https://arxiv.org/abs/hep-lat/9801021}
  {arXiv:hep-lat/9801021} \BibitemShut {NoStop}%
\bibitem [{\citenamefont {Kaplan}(1992)}]{Kaplan:1992bt}%
  \BibitemOpen
  \bibfield  {author} {\bibinfo {author} {\bibfnamefont {D.~B.}\ \bibnamefont
  {Kaplan}},\ }\href {https://doi.org/10.1016/0370-2693(92)91112-M} {\bibfield
  {journal} {\bibinfo  {journal} {Phys. Lett. B}\ }\textbf {\bibinfo {volume}
  {288}},\ \bibinfo {pages} {342} (\bibinfo {year} {1992})},\ \Eprint
  {https://arxiv.org/abs/hep-lat/9206013} {arXiv:hep-lat/9206013} \BibitemShut
  {NoStop}%
\bibitem [{\citenamefont {Shamir}(1993)}]{Shamir:1993zy}%
  \BibitemOpen
  \bibfield  {author} {\bibinfo {author} {\bibfnamefont {Y.}~\bibnamefont
  {Shamir}},\ }\href {https://doi.org/10.1016/0550-3213(93)90162-I} {\bibfield
  {journal} {\bibinfo  {journal} {Nucl. Phys. B}\ }\textbf {\bibinfo {volume}
  {406}},\ \bibinfo {pages} {90} (\bibinfo {year} {1993})},\ \Eprint
  {https://arxiv.org/abs/hep-lat/9303005} {arXiv:hep-lat/9303005} \BibitemShut
  {NoStop}%
\bibitem [{\citenamefont {Narayanan}\ and\ \citenamefont
  {Neuberger}(1994)}]{Narayanan:1993sk}%
  \BibitemOpen
  \bibfield  {author} {\bibinfo {author} {\bibfnamefont {R.}~\bibnamefont
  {Narayanan}}\ and\ \bibinfo {author} {\bibfnamefont {H.}~\bibnamefont
  {Neuberger}},\ }\href {https://doi.org/10.1016/0550-3213(94)90393-X}
  {\bibfield  {journal} {\bibinfo  {journal} {Nucl. Phys. B}\ }\textbf
  {\bibinfo {volume} {412}},\ \bibinfo {pages} {574} (\bibinfo {year}
  {1994})},\ \Eprint {https://arxiv.org/abs/hep-lat/9307006}
  {arXiv:hep-lat/9307006} \BibitemShut {NoStop}%
\bibitem [{\citenamefont {Narayanan}\ and\ \citenamefont
  {Neuberger}(1993)}]{Narayanan:1993ss}%
  \BibitemOpen
  \bibfield  {author} {\bibinfo {author} {\bibfnamefont {R.}~\bibnamefont
  {Narayanan}}\ and\ \bibinfo {author} {\bibfnamefont {H.}~\bibnamefont
  {Neuberger}},\ }\href {https://doi.org/10.1103/PhysRevLett.71.3251}
  {\bibfield  {journal} {\bibinfo  {journal} {Phys. Rev. Lett.}\ }\textbf
  {\bibinfo {volume} {71}},\ \bibinfo {pages} {3251} (\bibinfo {year}
  {1993})},\ \Eprint {https://arxiv.org/abs/hep-lat/9308011}
  {arXiv:hep-lat/9308011} \BibitemShut {NoStop}%
\bibitem [{\citenamefont {Neuberger}(1998{\natexlab{a}})}]{Neuberger:1997fp}%
  \BibitemOpen
  \bibfield  {author} {\bibinfo {author} {\bibfnamefont {H.}~\bibnamefont
  {Neuberger}},\ }\href {https://doi.org/10.1016/S0370-2693(97)01368-3}
  {\bibfield  {journal} {\bibinfo  {journal} {Phys. Lett. B}\ }\textbf
  {\bibinfo {volume} {417}},\ \bibinfo {pages} {141} (\bibinfo {year}
  {1998}{\natexlab{a}})},\ \Eprint {https://arxiv.org/abs/hep-lat/9707022}
  {arXiv:hep-lat/9707022} \BibitemShut {NoStop}%
\bibitem [{\citenamefont {Neuberger}(1998{\natexlab{b}})}]{Neuberger:1998wv}%
  \BibitemOpen
  \bibfield  {author} {\bibinfo {author} {\bibfnamefont {H.}~\bibnamefont
  {Neuberger}},\ }\href {https://doi.org/10.1016/S0370-2693(98)00355-4}
  {\bibfield  {journal} {\bibinfo  {journal} {Phys. Lett. B}\ }\textbf
  {\bibinfo {volume} {427}},\ \bibinfo {pages} {353} (\bibinfo {year}
  {1998}{\natexlab{b}})},\ \Eprint {https://arxiv.org/abs/hep-lat/9801031}
  {arXiv:hep-lat/9801031} \BibitemShut {NoStop}%
\bibitem [{\citenamefont {Azcoiti}\ \emph {et~al.}(2010)\citenamefont
  {Azcoiti}, \citenamefont {Di~Carlo}, \citenamefont {Follana},\ and\
  \citenamefont {Vaquero}}]{Azcoiti:2010ns}%
  \BibitemOpen
  \bibfield  {author} {\bibinfo {author} {\bibfnamefont {V.}~\bibnamefont
  {Azcoiti}}, \bibinfo {author} {\bibfnamefont {G.}~\bibnamefont {Di~Carlo}},
  \bibinfo {author} {\bibfnamefont {E.}~\bibnamefont {Follana}},\ and\ \bibinfo
  {author} {\bibfnamefont {A.}~\bibnamefont {Vaquero}},\ }\href
  {https://doi.org/10.1007/JHEP07(2010)047} {\bibfield  {journal} {\bibinfo
  {journal} {J. High Energy Phys.}\ }\textbf {\bibinfo {volume} {07}},\
  \bibinfo {pages} {047}},\ \Eprint {https://arxiv.org/abs/1004.3463}
  {arXiv:1004.3463 [hep-lat]} \BibitemShut {NoStop}%
\bibitem [{\citenamefont {Azcoiti}\ \emph {et~al.}(1995)\citenamefont
  {Azcoiti}, \citenamefont {Laliena},\ and\ \citenamefont
  {Luo}}]{Azcoiti:1995dq}%
  \BibitemOpen
  \bibfield  {author} {\bibinfo {author} {\bibfnamefont {V.}~\bibnamefont
  {Azcoiti}}, \bibinfo {author} {\bibfnamefont {V.}~\bibnamefont {Laliena}},\
  and\ \bibinfo {author} {\bibfnamefont {X.-Q.}\ \bibnamefont {Luo}},\ }\href
  {https://doi.org/10.1016/0370-2693(95)00602-H} {\bibfield  {journal}
  {\bibinfo  {journal} {Phys. Lett. B}\ }\textbf {\bibinfo {volume} {354}},\
  \bibinfo {pages} {111} (\bibinfo {year} {1995})},\ \Eprint
  {https://arxiv.org/abs/hep-th/9509091} {arXiv:hep-th/9509091} \BibitemShut
  {NoStop}%
\bibitem [{\citenamefont {Azcoiti}\ \emph {et~al.}(2008)\citenamefont
  {Azcoiti}, \citenamefont {di~Carlo},\ and\ \citenamefont
  {Vaquero}}]{Azcoiti:2008nq}%
  \BibitemOpen
  \bibfield  {author} {\bibinfo {author} {\bibfnamefont {V.}~\bibnamefont
  {Azcoiti}}, \bibinfo {author} {\bibfnamefont {G.}~\bibnamefont {di~Carlo}},\
  and\ \bibinfo {author} {\bibfnamefont {A.}~\bibnamefont {Vaquero}},\ }\href
  {https://doi.org/10.1088/1126-6708/2008/04/035} {\bibfield  {journal}
  {\bibinfo  {journal} {J. High Energy Phys.}\ }\textbf {\bibinfo {volume}
  {04}},\ \bibinfo {pages} {035}},\ \Eprint {https://arxiv.org/abs/0804.1338}
  {arXiv:0804.1338 [hep-th]} \BibitemShut {NoStop}%
\bibitem [{Note1()}]{Note1}%
  \BibitemOpen
  \bibinfo {note} {I refer here to the exact flavor symmetry of several
  staggered fields with the same mass, that holds at any lattice spacing, not
  to the approximate taste symmetry that becomes exact only in the continuum
  limit.}\BibitemShut {Stop}%
\bibitem [{Note2()}]{Note2}%
  \BibitemOpen
  \bibinfo {note} {The result applies to $\gamma _5$-Hermitean Ginsparg-Wilson
  fermions with $2R=\protect \mathbf {1}$, see Section \ref
  {sec:gwf}.}\BibitemShut {Stop}%
\bibitem [{\citenamefont {Mermin}\ and\ \citenamefont
  {Wagner}(1966)}]{Mermin:1966fe}%
  \BibitemOpen
  \bibfield  {author} {\bibinfo {author} {\bibfnamefont {N.~D.}\ \bibnamefont
  {Mermin}}\ and\ \bibinfo {author} {\bibfnamefont {H.}~\bibnamefont
  {Wagner}},\ }\href {https://doi.org/10.1103/PhysRevLett.17.1133} {\bibfield
  {journal} {\bibinfo  {journal} {Phys. Rev. Lett.}\ }\textbf {\bibinfo
  {volume} {17}},\ \bibinfo {pages} {1133} (\bibinfo {year}
  {1966})}\BibitemShut {NoStop}%
\bibitem [{\citenamefont {Hohenberg}(1967)}]{Hohenberg:1967zz}%
  \BibitemOpen
  \bibfield  {author} {\bibinfo {author} {\bibfnamefont {P.~C.}\ \bibnamefont
  {Hohenberg}},\ }\href {https://doi.org/10.1103/PhysRev.158.383} {\bibfield
  {journal} {\bibinfo  {journal} {Phys. Rev.}\ }\textbf {\bibinfo {volume}
  {158}},\ \bibinfo {pages} {383} (\bibinfo {year} {1967})}\BibitemShut
  {NoStop}%
\bibitem [{\citenamefont {Coleman}(1973)}]{Coleman:1973ci}%
  \BibitemOpen
  \bibfield  {author} {\bibinfo {author} {\bibfnamefont {S.~R.}\ \bibnamefont
  {Coleman}},\ }\href {https://doi.org/10.1007/BF01646487} {\bibfield
  {journal} {\bibinfo  {journal} {Commun. Math. Phys.}\ }\textbf {\bibinfo
  {volume} {31}},\ \bibinfo {pages} {259} (\bibinfo {year} {1973})}\BibitemShut
  {NoStop}%
\bibitem [{Note3()}]{Note3}%
  \BibitemOpen
  \bibinfo {note} {On a finite lattice observables are necessarily polynomial
  in the fermion fields and dependent on finitely many link variables: the
  restriction is relevant in the thermodynamic limit. Nonzero fermion number
  immediately entails a vanishing expectation value. Any observable that is not
  gauge-invariant can be replaced with its average over gauge orbits without
  changing its expectation value.}\BibitemShut {Stop}%
\bibitem [{\citenamefont {Naik}(1989)}]{Naik:1986bn}%
  \BibitemOpen
  \bibfield  {author} {\bibinfo {author} {\bibfnamefont {S.}~\bibnamefont
  {Naik}},\ }\href {https://doi.org/10.1016/0550-3213(89)90394-5} {\bibfield
  {journal} {\bibinfo  {journal} {Nucl. Phys. B}\ }\textbf {\bibinfo {volume}
  {316}},\ \bibinfo {pages} {238} (\bibinfo {year} {1989})}\BibitemShut
  {NoStop}%
\bibitem [{\citenamefont {Lepage}(1999)}]{Lepage:1998vj}%
  \BibitemOpen
  \bibfield  {author} {\bibinfo {author} {\bibfnamefont {G.~P.}\ \bibnamefont
  {Lepage}},\ }\href {https://doi.org/10.1103/PhysRevD.59.074502} {\bibfield
  {journal} {\bibinfo  {journal} {Phys. Rev. D}\ }\textbf {\bibinfo {volume}
  {59}},\ \bibinfo {pages} {074502} (\bibinfo {year} {1999})},\ \Eprint
  {https://arxiv.org/abs/hep-lat/9809157} {arXiv:hep-lat/9809157} \BibitemShut
  {NoStop}%
\bibitem [{\citenamefont {Morningstar}\ and\ \citenamefont
  {Peardon}(2004)}]{Morningstar:2003gk}%
  \BibitemOpen
  \bibfield  {author} {\bibinfo {author} {\bibfnamefont {C.}~\bibnamefont
  {Morningstar}}\ and\ \bibinfo {author} {\bibfnamefont {M.~J.}\ \bibnamefont
  {Peardon}},\ }\href {https://doi.org/10.1103/PhysRevD.69.054501} {\bibfield
  {journal} {\bibinfo  {journal} {Phys. Rev. D}\ }\textbf {\bibinfo {volume}
  {69}},\ \bibinfo {pages} {054501} (\bibinfo {year} {2004})},\ \Eprint
  {https://arxiv.org/abs/hep-lat/0311018} {arXiv:hep-lat/0311018} \BibitemShut
  {NoStop}%
\bibitem [{\citenamefont {Follana}\ \emph {et~al.}(2007)\citenamefont
  {Follana}, \citenamefont {Mason}, \citenamefont {Davies}, \citenamefont
  {Hornbostel}, \citenamefont {Lepage}, \citenamefont {Shigemitsu},
  \citenamefont {Trottier},\ and\ \citenamefont {Wong}}]{Follana:2006rc}%
  \BibitemOpen
  \bibfield  {author} {\bibinfo {author} {\bibfnamefont {E.}~\bibnamefont
  {Follana}}, \bibinfo {author} {\bibfnamefont {Q.}~\bibnamefont {Mason}},
  \bibinfo {author} {\bibfnamefont {C.}~\bibnamefont {Davies}}, \bibinfo
  {author} {\bibfnamefont {K.}~\bibnamefont {Hornbostel}}, \bibinfo {author}
  {\bibfnamefont {G.~P.}\ \bibnamefont {Lepage}}, \bibinfo {author}
  {\bibfnamefont {J.}~\bibnamefont {Shigemitsu}}, \bibinfo {author}
  {\bibfnamefont {H.}~\bibnamefont {Trottier}},\ and\ \bibinfo {author}
  {\bibfnamefont {K.}~\bibnamefont {Wong}} (\bibinfo {collaboration} {HPQCD,
  UKQCD Collaborations}),\ }\href {https://doi.org/10.1103/PhysRevD.75.054502}
  {\bibfield  {journal} {\bibinfo  {journal} {Phys. Rev. D}\ }\textbf {\bibinfo
  {volume} {75}},\ \bibinfo {pages} {054502} (\bibinfo {year} {2007})},\
  \Eprint {https://arxiv.org/abs/hep-lat/0610092} {arXiv:hep-lat/0610092}
  \BibitemShut {NoStop}%
\bibitem [{\citenamefont {Wilson}(1975)}]{Wilson:1975id}%
  \BibitemOpen
  \bibfield  {author} {\bibinfo {author} {\bibfnamefont {K.~G.}\ \bibnamefont
  {Wilson}},\ }in\ \href@noop {} {\emph {\bibinfo {booktitle} {{13th
  International School of Subnuclear Physics: New Phenomena in Subnuclear
  Physics}}}}\ (\bibinfo {year} {1975})\BibitemShut {NoStop}%
\bibitem [{Note4()}]{Note4}%
  \BibitemOpen
  \bibinfo {note} {Proof: In matrix notation, Eq.~\protect \textup {\hbox
  {\mathsurround \z@ \protect \normalfont (\ignorespaces \ref
  {eq:alt_irrep}\unskip \@@italiccorr )}} reads $u^{(R)T}= u^{(R)T}D^{(R)}(V)$,
  $\forall V\in \protect \mathrm {SU}(N_f)$, implying also
  $u^{(R)*}=D^{(R)}(V)^\protect \dag u^{(R)*}$. The projector
  $P=u^{(R)*}u^{(R)T}$ is then left invariant by an irreducible unitary
  representation of $\protect \mathrm {SU}(N_f)$, $P = D^{(R)}(V)^\protect \dag
  P D^{(R)}(V)= D^{(R)}(V)^{-1} P D^{(R)}(V)$. By Schur's lemma $P$ is then
  proportional to the $d_R$-dimensional identity matrix, but since it has only
  rank 1 it must vanish if $d_R\not =1$.}\BibitemShut {Stop}%
\end{thebibliography}%

\end{document}